\documentclass[useAMS,usenatbib,usegraphicx]{mn2e}

\usepackage{epsfig}

\newcommand{\HI}{\mbox{H\,{\sc i}}}
\newcommand{\Lya}{Lyman-$\alpha$}
\newcommand{\Lyb}{Lyman-$\beta$}
\newcommand{\AlII}{\mbox{Al\,{\sc ii}}}
\newcommand{\AlIII}{\mbox{Al\,{\sc iii}}}
\newcommand{\CII}{\mbox{C\,{\sc ii}}}
\newcommand{\CIV}{\mbox{C\,{\sc iv}}}
\newcommand{\FeII}{\mbox{Fe\,{\sc ii}}}

\newcommand{\MgII}{\mbox{Mg\,{\sc ii}}}

\newcommand{\SiII}{\mbox{Si\,{\sc ii}}}
\newcommand{\SiIII}{\mbox{Si\,{\sc iii}}}
\newcommand{\SiIV}{\mbox{Si\,{\sc iv}}}
\newcommand{\OVI}{\mbox{O\,{\sc vi}}}

\newcommand{\kms}{km~s$^{-1}$}

\def\pa{{\parallel}}
\def\pe{{\perp}}
\def\los{line of sight}
\def\loss{lines of sight}

\title[High-resolution spectroscopy of the 3D cosmic web]{ 
  High resolution spectroscopy of the three dimensional cosmic web 
  with close QSO groups
  \thanks{Based on observations collected at the
  European Southern Observatory Very Large Telescope, Cerro Paranal,
  Chile -- Programs  68.A-0216(A), 69.A-0204(A),
  69.A-0586(A), 70.A-0031(A), 077.A-0714(A) }}

  \author[M. Cappetta et al.]{M. Cappetta$^{1,2}$\thanks{E-mail: cappetta@mpe.mpg.de}, 
  V. D'Odorico$^{2}$, S. Cristiani$^{2,3}$, F. Saitta$^{2}$, M. Viel$^{2,3}$\\   
$^1$MPE -- Max-Planck-Institut f\"{u}r Extraterrestrische Physik, Giessenbachstrasse 1, 85748 Garching bei M\"{u}nchen, Germany\\
$^2$INAF -- Osservatorio Astronomico di Trieste, via G.B. Tiepolo 11, I-34127 Trieste, Italy\\ 
$^3$INFN/National Institute for Nuclear Physics, Via Valerio 2, I-34127 Trieste, Italy\\}

\begin{document}

\date{Accepted 2010 May 8. Received 2010 May 6; in original form 2009 October 13}

\pagerange{\pageref{firstpage}--\pageref{lastpage}} \pubyear{2009}

\maketitle

\label{firstpage}

\begin{abstract} 
We study the three-dimensional distribution of matter at $z\sim2$
using high resolution spectra of QSO pairs and
simulated spectra drawn from cosmological hydro-dynamical
simulations. We present a sample of 15 QSOs, corresponding 
to 21 baselines of angular separations evenly distributed
between $\sim 1$ and 14 arcmin, observed with 
the Ultraviolet and Visual Echelle Spectrograph (UVES) at
the European Southern Observatory-Very Large Telescope (ESO-VLT).
The observed correlation functions of the transmitted flux in the 
\HI\ \Lya\ forest transverse to and along the \los\ are in agreement, implying
that the distortions in redshift space due to peculiar velocities are
relatively small and - within the relatively large error bars - not significant. 
The clustering signal is significant up to velocity
separations of $\sim 300$ \kms, corresponding to about 5 $h^{-1}$ comoving Mpc. 
Compatibility at the $2\ \sigma$ level has been found both for the
Auto- and Cross-correlation functions and for the set of the
Cross correlation coefficients. The analysis focuses in particular on two
QSO groups of the sample, the Sextet and the Triplet. Searching for alignments in the redshift
space between \Lya\ absorption lines belonging to different \loss, it
has been possible to discover the presence of a wide \HI\
structures extending over about ten Mpc in comoving space, and give
constraints on the sizes of two cosmic under-dense regions in the
intergalactic medium, which have been detected with a 91\% and 86\% significance level, 
respectively in the Sextet and in the Triplet.
\end{abstract}


\section{Introduction} 

The understanding of the \Lya\ forest has dramatically
improved in the recent decade, both on the theoretical and the
observational side. Semi-analytical and hydro-dynamical simulations
have outlined a new picture where the \Lya\ forest is due to
the fluctuations of the intermediate and low-density intergalactic
medium (IGM), arising naturally in the hierarchical process of
structure formation. Relatively simple physical processes impact on the
thermal state of the gas, which, on scales larger than the Jeans
length, effectively traces the underlying distribution of dark matter.
Support for this scenario is given by the satisfactory reproduction by 
semi-analytical and hydro-simulations of many properties of the \Lya\ forest
(from the column density and the Doppler parameter distribution to the
number density and effective opacity evolution) derived from the
analysis of high resolution, high signal-to-noise ratio (SNR) QSO spectra
obtained at 8-10m class telescopes 
\citep[e.g.,][]{dave99,brianmachacek00,kim01,bianchi03,janknecht06}.
A step forward in the study of the IGM with QSO absorption spectra is
represented by the use of multiple \loss\ at close
angular separations,  which allows information about the
transverse direction to be obtained. Common absorption features observed in the
spectra of multiply lensed quasars \citep[e.g.,][]{smette92} and of
close quasar pairs \citep[e.g.,][]{vale98,aracil02}
have provided evidence that the \Lya\ absorbers have
dimensions of a few hundred kpc, in agreement with the predictions of
simulations. Recently, lensed and more widely separated QSO pairs have
been used to recover the kinematics of the gaseous cosmic web \citep{rauch05},
 confirming that the Hubble expansion and gravitational
instability are the main processes influencing the \Lya\ forest gas.

A critical test of the nature of the \Lya\ absorbers, as proposed by 
simulations, comes from the determination of their spatial distribution 
properties. This has been done by analysing a great 
number of uncorrelated QSO lines of sight and computing the flux correlation
function by averaging over many spectra \citep[e.g.,][]{tripp98,savaglio99,croft02}.
In this observational approach, however, 
the three-dimensional information is convolved with distortions in
redshift space, due to peculiar motions and thermal
broadening. Multiple \loss\ at small angular separations offer an
invaluable alternative to address the spatial  
distribution of the absorbers, enabling a more direct interpretation of 
the observed correlations. 

The final goal of this work is to investigate the distribution properties 
of matter in the IGM applying the modern interpretation of the
\Lya\ forest to a sample of close QSO groups.
The computation and comparison of the flux correlation function along
and across the lines of sight was carried out in a previous paper
\citep[][Paper I]{vale06}. In this paper, we update the previous
results using higher SNR spectra and, in particular,
we investigate the coincidences of absorbers among three or more close lines
of sight in order to detect cosmological structures extending to large
scales. Finally, we look for extended under-dense regions in the IGM
using multiple lines of sight. 

Note that tomographic studies based on Lyman-alpha lines similar to those 
presented here will be of great importance in the near future due to the 
large numbers of quasar spectra that will be collected using low and medium 
resolution spectrographs (BOSS survey and the X-Shooter instrument for example) 
that aim at constraining spatial matter correlations in the transverse and 
longitudinal directions to ultimately measure baryonic acoustic oscillations 
and perform the Alcock-Paczyinski test at high redshift.

The paper structure is the following: in Section 2 we describe the
observed data sample, the reduction procedure and the simulated
spectra.  
Section 3 is devoted to the computation of the auto- and
cross-correlation functions and the cross correlation coefficients of
both the observed and simulated spectra and to their comparison. In
Section 4, we analyze the coincidences in redshift space of the
\HI\ and \CIV\ absorbers in order to detect structures extending
over several comoving Mpc, while the under-dense regions common to 
multiple lines of sight are studied in Section 5. The conclusions 
are then drawn in Section 6. 
Throughout this paper we adopt $\Omega_{0\rm{m}} = 0.26,\
  \Omega_{0\Lambda}=0.74$, and  $h=H_0/(72$ km s$^{-1}$ Mpc$^{-1})$.

\begin{table}
  \centering
    \caption{Characteristics of the observed QSO spectra.}
    \label{obs:qso}
    \begin{tabular}{@{}rcccc@{}}
      \hline 
    Object & $z$ & M$_{\rm B}$ & \Lya\ & SNR  \\ 
& & & range & per pixel \\
\hline 
Pair A\ \ \ \ \ \ \ \ \  PA1 & 2.645 & 19.11 & 2.094--2.585 & 8--12 \\
\ \ \ \ \ \ \ \ \ \ \ \ \ \ PA2 & 2.610 & 19.84 & 2.094--2.550 & 3.5--6.5 \\
\ \ \ \ \ \ \ \ \ \ \ \ \ \   &   &   &   &   \\
Triplet\ \ \ \ \ \ \ \ \ \ \ T1 &  2.041 & 18.20 & 1.633--1.991 & 3--15 \\
       \ \ \ \ \ \ \ \ \ \ \ T2 &  2.05  & 18.30 & 1.592--1.999 & 4--15 \\
       \ \ \ \ \ \ \ \ \ \ \ T3 &  2.053 & 18.10 & 1.665-2.002 & 2.5--7 \\
	   \ \ \ \ \ \ \ \ \ \ \ \ \ \   &   &   &   &   \\
Sextet\ \ \ \ \ \ \ \ \ \ \ \ S1 &  1.907 & 19.66 & 1.665--1.859 & 2--7 \\
      \ \ \ \ \ \ \ \ \ \ \ \ S2 &  2.387 & 19.53 & 1.858--2.331 & 3--8\\
      \ \ \ \ \ \ \ \ \ \ \ \ S3 &  2.102 & 19.31 & 1.633--2.051 & 4--12 \\
      \ \ \ \ \ \ \ \ \ \ \ \ S4 &  1.849 & 19.59 & 1.575--1.802 & 3--9 \\
      \ \ \ \ \ \ \ \ \ \ \ \ S5 &  2.121 & 18.85 & 1.633--2.069 & 4--12\\
      \ \ \ \ \ \ \ \ \ \ \ \ S6 &  2.068 & 20.19 & 1.592--2.017 & 3--10 \\
	  \ \ \ \ \ \ \ \ \ \ \ \ \ \   &   &   &   &   \\
Pair U\ \ \ \ \ UM680  & 2.144 & 18.60 & 1.653--2.092 & 6.5--17 \\
      \ \ \ \ \ UM681  & 2.122 & 19.10 & 1.634--2.070 & 7--17 \\
	  \ \ \ \ \ \ \ \ \ \ \ \ \ \   &   &   &   &   \\
Pair Q Q2343+12 & 2.549 &17.00 & 1.994--2.490 & 13--23 \\
       Q2344+12 & 2.773 & 17.50 & 2.183--2.711 & 12--18 \\
\hline \\
\end{tabular}
\end{table}

\section{Data sample}  

\begin{figure*}
\includegraphics[height=8truecm,width=8.5truecm]{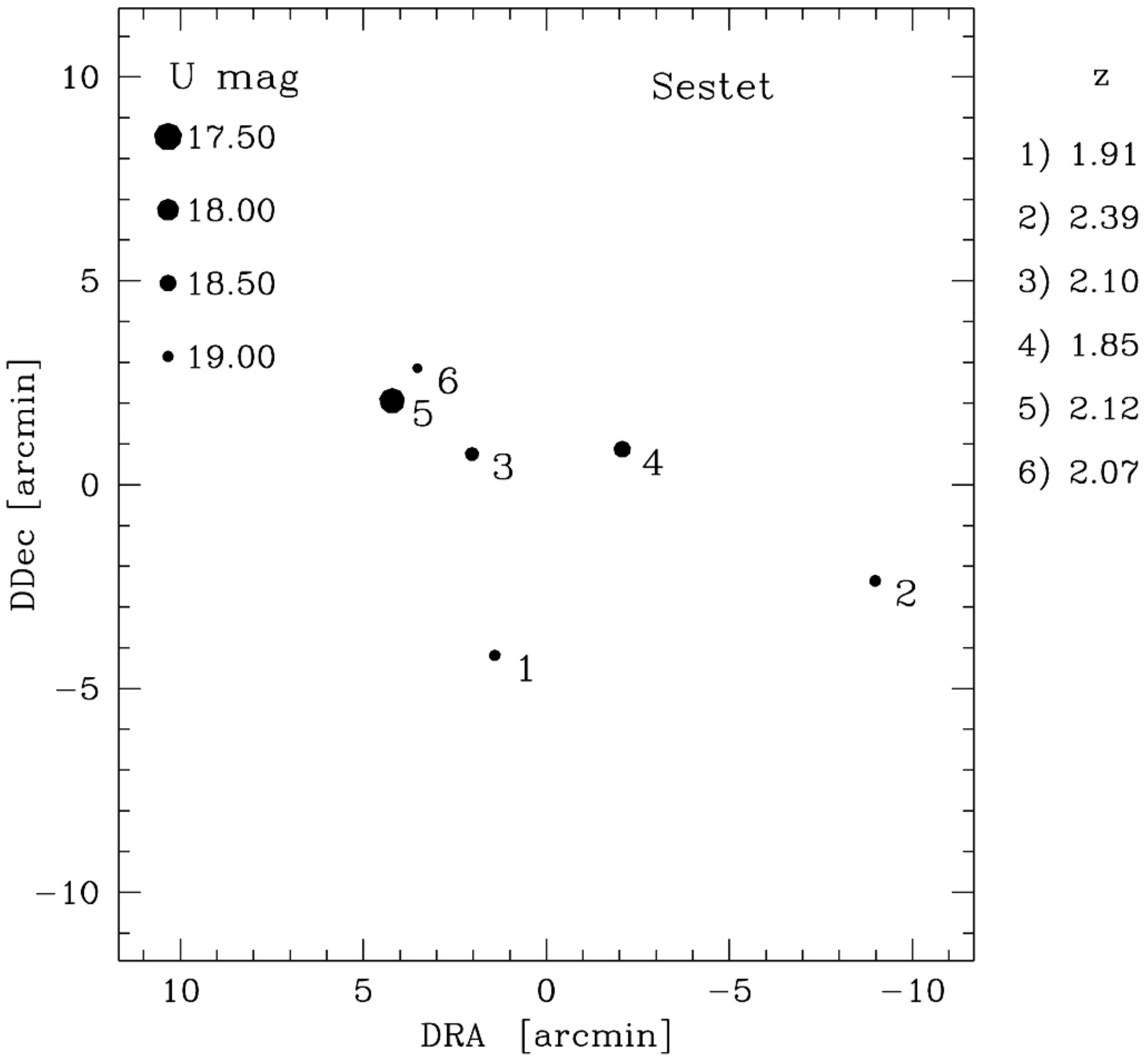}
\includegraphics[height=8truecm,width=8.5truecm]{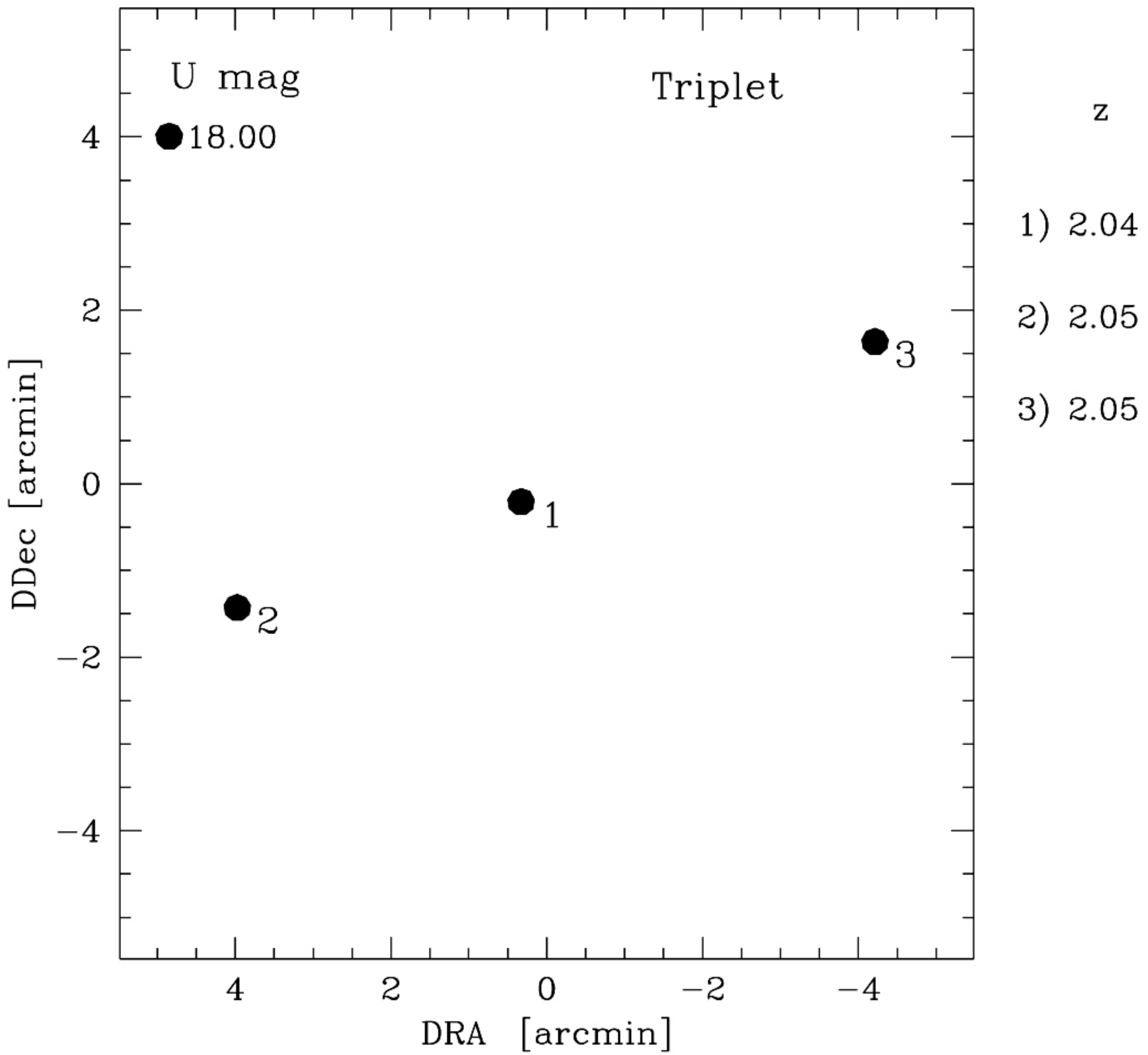}
  \caption{Relative positions, U magnitudes and redshifts of the QSOs
  composing the Sextet (left plot) and the Triplet (right plot) in our
  sample.}\label{fig:sextet} 
\end{figure*}

The QSO pairs and groups forming our sample are the same as Paper I 
and are described in Table~\ref{obs:qso}.  
The patterns on the sky of the two QSO groups called the Sextet and
the Triplet are reported in Fig.~\ref{fig:sextet}.

We were allocated 17 hours with UVES at the VLT in order to increase
the SNR of the spectra of objects T1 and T2 of
the Triplet and S1, S2, S5 and S6 of the Sextet. The new observations
were reduced with the UVES pipeline following a standard
procedure \citep{ballester00}.
A pre-filtering of the cosmic rays of the blue band frames was necessary for
the faintest spectra to carry out the optimal extraction properly. 

Then, the heliocentric and vacuum wavelength corrections were
performed and the new spectra were combined with the old ones.  
The final spectra have resolution $R \sim 45000$, while the SNR per
pixel varies on average between 3 and 12 in the \Lya\ forest and
between 6 and 20 in the  \CIV\ forest (see Tab.~\ref{obs:qso} for details). 

The estimate of the continuum level, in particular in the \Lya\ forest
region, is a very delicate step in the process of spectra
reduction.  Procedures realized up to now in order to determine 
the continuum position through automatic algorithms do not
give satisfactory results. Following the same procedure adopted in
Paper I, we fitted the regions free from clear absorption with a
spline polynomial of third order.  The limitations introduced by
the uncertainty in the true continuum level should play a minor role in
the computation of the cross-correlation function with respect to the 
the case of single line of sight analysis. This is because the power on
 scales of the continuum fluctuations is uncorrelated between the different lines of sight.
Thus, probing the transverse direction could potentially allow to
measure the matter distribution at scales larger than those probed along 
the \los\ \citep[see][for a discussion in the case of power 
spectra]{viel2002}.

 All the lines falling in the \Lya\ forest have been fitted with a Voigt 
profile via $\chi^{2}$ minimization. The lines with an equivalent
width (EW) lower than three times the related 
EW uncertainty have been removed from the list of fitted lines. 
The metal lines have been identified first looking for the most common
doublets (e.g. \CIV, \SiIV, \OVI, and \MgII). Then, we have searched for other
common transitions (e.g. \SiIII, \SiII, \CII, \FeII) at the redshift
of the previously determined systems.  

Our sample provides 21 QSO pairs with angular separations
uniformly distributed between $\sim 1$ and 14 arcmin, corresponding to
comoving spatial separations between $\sim 1.4$ and 21.6 h$^{-1}$ Mpc.  
The median redshift of the \Lya\ forest is $z \sim 1.8$. This is the
largest sample of high-resolution spectra of QSO pairs ever collected,
unique both for the number density - we have six QSOs in a region  
of $\sim 0.04$ deg$^{2}$ - and the variety of \los\ separations
investigated. 

\subsection{Simulated spectra}

In order to both assess the nature of the \Lya\ forest inferred from 
simulations and to constrain the cosmological scenario of the same 
simulations, we compared the results obtained for our sample of 
observed QSO spectra with analogous results for a sample of mock 
\Lya\ forests. 

The details of the adopted simulations can be found in Paper I,
here we provide only the basic information. 
We used simulations run with the parallel hydro-dynamical (TreeSPH)
code {\small {GADGET-2}} \citep{springel2001,springel2005}. The
simulations were performed with periodic boundary conditions with an
equal number of dark matter and gas particles and used the
conservative `entropy-formulation' of SPH proposed by
\citet{springel2002}. 
Radiative cooling and heating processes were followed for a primordial
mix of hydrogen and helium. We assumed a mean UV background
produced by quasars and galaxies as given by \citet{hm96}
with helium heating rates multiplied by a factor 3.3 in order to fit
observational constraints on the temperature evolution of the IGM.  
More details can be found in \citet{viel2004}. 

The cosmological model corresponds to a `fiducial' $\Lambda$CDM
universe with parameters $\Omega_{0\rm{m}}=0.26,\
\Omega_{0\Lambda}=0.74,\ \Omega_{0\rm{b}}=0.0463$ and $H_0 = 72$ km
s$^{-1}$ Mpc$^{-1}$ (the B2 series of \citealt{viel2004}). 
We have used $2\times 400^3$ dark matter and
gas particles in a $120\ h^{-1}$ comoving Mpc box. 
The gravitational softening was set to 5 $h^{-1}$ kpc in
comoving units for all particles.
We note that the parameters chosen here, including the thermal
history of the IGM, are in perfect agreement with observational
constraints including recent results on the CMB and other results
obtained by the \Lya\ forest community
\citep[e.g. ][]{spergel03,viel2004,seljak05}.

The $z=1.8$ output of the simulated box was pierced by three \loss\ 
in order to obtain 50 triplets of spectra carefully reproducing 
the observed Triplet mutual separations and spectral properties.   
The same was done for 50 sextets of \loss\ reproducing the observed
Sextet and 50 pairs of \loss\ at the same angular separation as Pair U. 
50 different realizations of Pair A spectra and of Pair Q were
obtained from the output box at redshift $z=2.4$.    

Finally, we added to the simulated spectra both instrumental
broadening and a Gaussian noise in order to reproduce the observed
average SNR (per pixel): $SNR=5$ for the Triplet, the Sextet and Pair
A; $SNR=9$ for Pair U and $SNR=15$ for Pair Q.

\section{Correlation functions of the IGM}

In this section, we discuss the flux correlations in the absorption
spectra of our sample of QSO pairs. The statistical quantities are
the same as those already computed in Paper I, here we verify the 
effect of having an increased SNR.

On the basis of the interpretation of the \Lya\ forest 
as due to a continuous density field with a one-to-one correspondence between 
density and transmitted flux, we computed the correlation properties of the 
transmitted flux in QSO \loss\ and regarded them as indicators of the correlation 
properties of matter in the IGM.
We selected in each normalized spectrum the region between the
\Lyb\ emission (or the shortest observed wavelength, when the
\Lyb\ was not included in the spectrum) and 5000 \kms\ from the
\Lya\ emission (to avoid proximity effect due  
to the QSO). Absorption lines due to ions of elements heavier than hydrogen 
contaminate the \Lya\ forest and can give spurious contributions to the 
clustering signal \citep[see][for a discussion in the case of single 
lines of sight]{kim04}. We flagged and removed the spectral regions
where metal lines and \Lya\ absorptions of damped and sub-damped
systems occurred inside the \Lya\ forest.

Given the normalized transmitted flux, $f$, as a function of
the velocity $v_{\pa}$ along the \los\ and the angular position
$\theta$ on the sky, we define $\delta_f = (f - \bar{f})$,
where the average flux, $\bar{f}$, is computed for every spectrum as
the mean of the transmitted flux over all the considered 
pixels in that spectrum. 
We neglected the redshift evolution of the average transmitted
flux in the \Lya\ forest of the individual spectra, which translates
into the redshift evolution of the mean \HI\ opacity of the Universe
\citep{kim02,schaye03,viel2004,kirkman05,fg2008}, because we verified
that its effect  on the correlation function is negligible.
By means of this new field, $\delta_f$, we could then define and compute
three useful tools for the investigation of the \Lya\ correlation properties: 
the Auto and Cross correlation function and the set of the Cross 
correlation coefficients.

\subsection{The Auto-correlation function}

The unnormalized Auto correlation function (Auto CF) of the flux along the \los\
is defined as: 

\begin{equation}
\xi^f_{\pa}(\Delta v_{\pa}) = \langle \delta_f(v_{\pa})
\delta_f(v_{\pa}+\Delta v_{\pa})\rangle,  
\end{equation}

\par\noindent
following previous studies on the same subject \citep[e.g.][]{mcdonald00,rollinde03,becker04}.
The Auto CF for our sample of QSO spectra was obtained by averaging over all 
the pixels of all the QSOs. The results were binned in 50 \kms\ velocity bins. 
The Auto CF for the simulated spectra was computed as the arithmetic mean of 
the correlation functions obtained for 50 realizations of the observed sample 
and the error is the corresponding standard deviation. It is important to 
recognize that the computed error bars for the simulated  $\xi^f_{\pa}$ are 
strongly correlated. This is due to the fact that every pixel contributes to 
the correlation function in several velocity bins. 

\begin{figure}
\includegraphics[height=7truecm,width=8truecm]{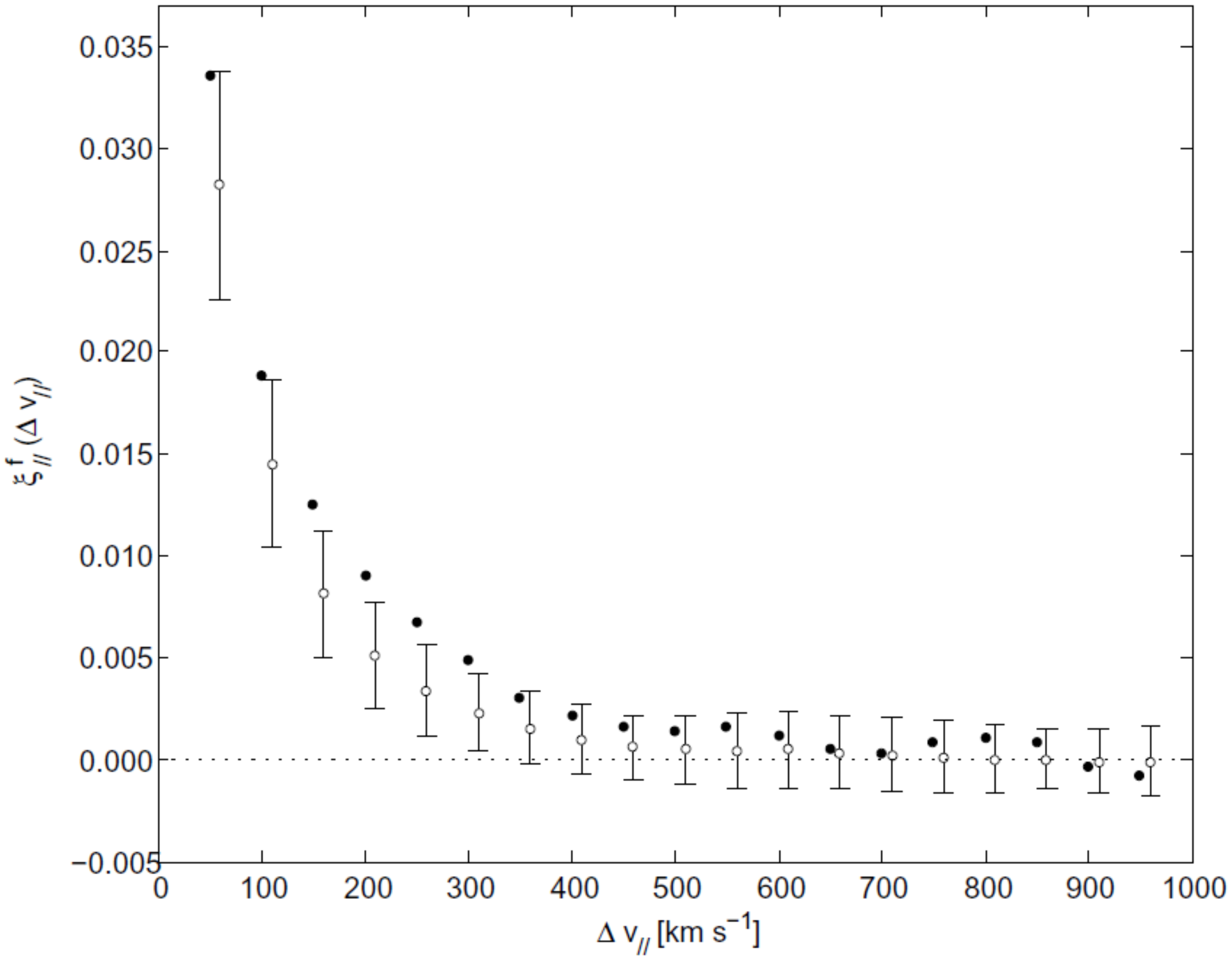}
  \caption{Comparison of the observed and simulated Auto-correlation functions: 
  solid dots refer to the observations and empty dots 
  to the simulations. Simulated data are slightly shifted in 
  velocity for clarity.}\label{fig:auto_CF} 
\end{figure}

In Fig.~\ref{fig:auto_CF} we show the comparison of the observed and 
simulated Auto CF in velocity space.  
With the improved SNR with respect to Paper I, the agreement between
the two correlation functions has increased, weakening any evidence for
extra clustering in the regions occupied by QSO pairs. 

\subsection{The Cross-correlation function}

In this section we exploit the capabilities of our sample of QSO pairs by 
determining the clustering properties of the IGM across the \loss. The great 
advantage with respect to the correlation function along the \los, in particular 
for a sample like ours showing a large variety of pair separations, is that we 
have the guarantee of sampling true spatial separations between the pixels, 
the effect of peculiar velocities being negligible or absent. As a first approach, 
we computed the Cross-correlation function (Cross CF) extending in a natural way 
the procedure adopted for the Auto CF. 

Every pixel along the \los\ is considered as an element of the density
field at the QSO angular position in the sky and at a distance from
the observer (comoving along the \los) corresponding to the 
redshift of the pixel:

\begin{equation}
r_{\pa}(z) = \frac{c}{H_0} \int_0^z \frac{{\rm d}z'}{E(z')},  
\end{equation}

\noindent
where $E(z)$ describes the evolution of the Hubble parameter as a function 
of the redshift\footnote{$E(z)  = \sqrt{\Omega_{0\rm{m}}\,(1+z)^3 + \Omega_{0\Lambda}}$}.

In the definition of the comoving distance there is the implicit hypothesis 
that peculiar velocities give a negligible contribution to the
measured redshift in the \Lya\ forest \citep[][Paper I]{rauch05}.
The Cross CF of the transmitted flux between two lines of sight at angular 
separation $\Delta\theta$ is defined as

\begin{equation}
\xi^f_{\times}(\Delta r) =
\langle\delta_f(\theta,r_{\pa,1})
\delta_f(\theta+\Delta\theta,r_{\pa,2})\rangle,  
\end{equation}

\noindent
where, $\Delta r = \sqrt{(r_{\pa,1}^2 + r_{\pa,2}^2 - 2\,
r_{\pa,1}\,r_{\pa,2}\,\cos\Delta\theta)}$ is the spatial separation
between pixel 1 at $r_{\pa,1}$ along one \los\ and pixel 2 at
$r_{\pa,2}$ along the paired \los.

\begin{figure}
\includegraphics[height=7truecm,width=8truecm]{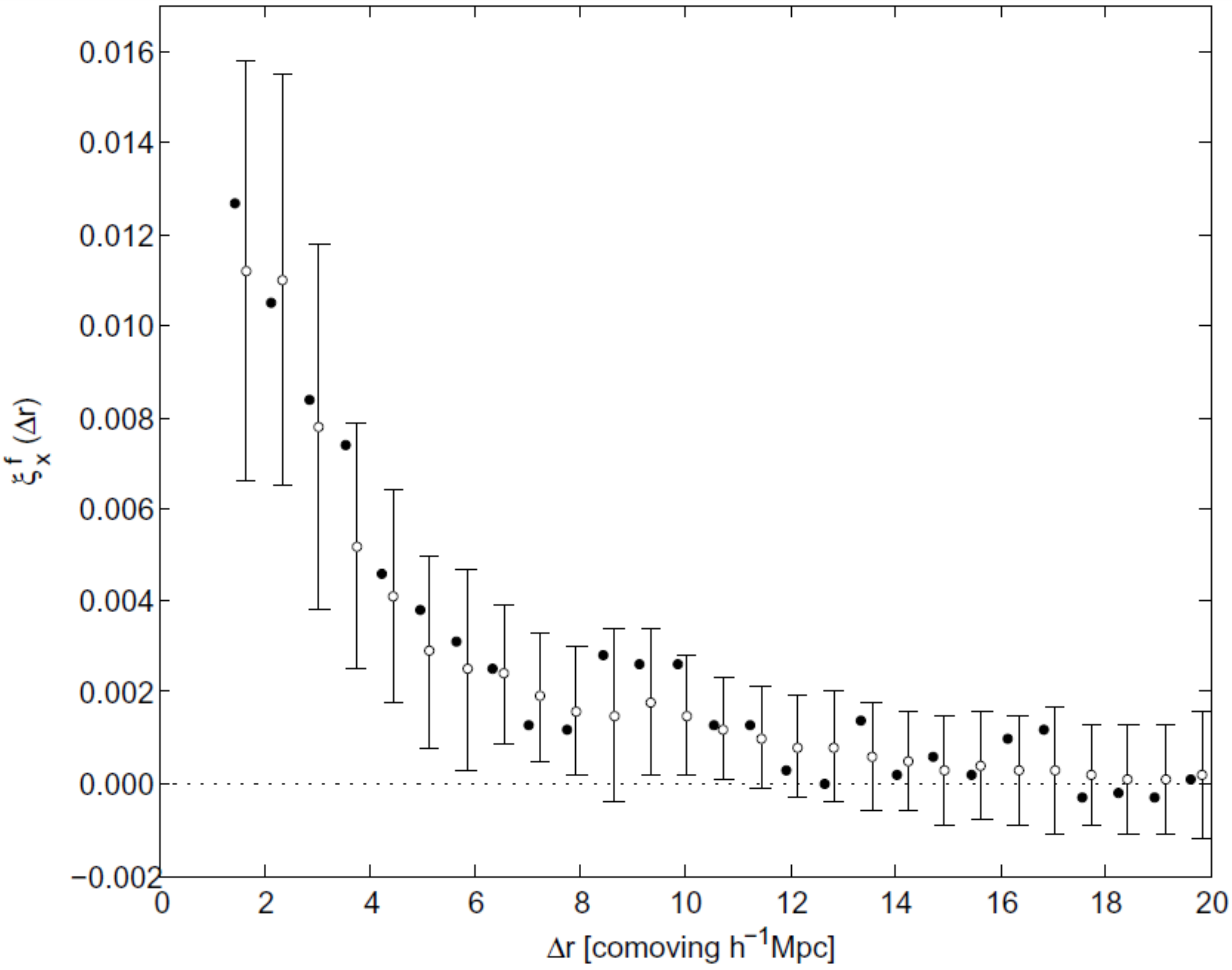}
  \caption{Comparison of Cross-correlation functions: solid dots are
    related to the observations and empty points to the simulations. Simulations 
	data are slightly shifted in comoving distance for clarity.}\label{fig:cross_CF} 
\end{figure}

We also computed the Cross CF for the sample of mock spectra. 
The simulated spectra are characterized by the redshift of the output box and 
a velocity extent. In order to assign a redshift value to 
every pixel, we gave the central pixel of every spectrum the redshift of the 
corresponding output box, then we numbered the pixels one by one, transforming 
the velocity size of the pixel into a redshift size. Once the redshifts were 
determined pixel by pixel, we followed the same procedure adopted for the observed 
spectra for the 50 simulated samples and computed the average Cross CF and its 
1~$\sigma$ standard deviation. The result of our computation is shown in 
Fig.~\ref{fig:cross_CF} compared with the observed Cross CF 
of the pairs. 
The two functions show very good agreement (at the 1~$\sigma$ level) and a 
significant clustering signal up to comoving separations of $\sim  4\ h^{-1}$ comoving Mpc.
 
\begin{figure}
\includegraphics[height=7truecm,width=8truecm]{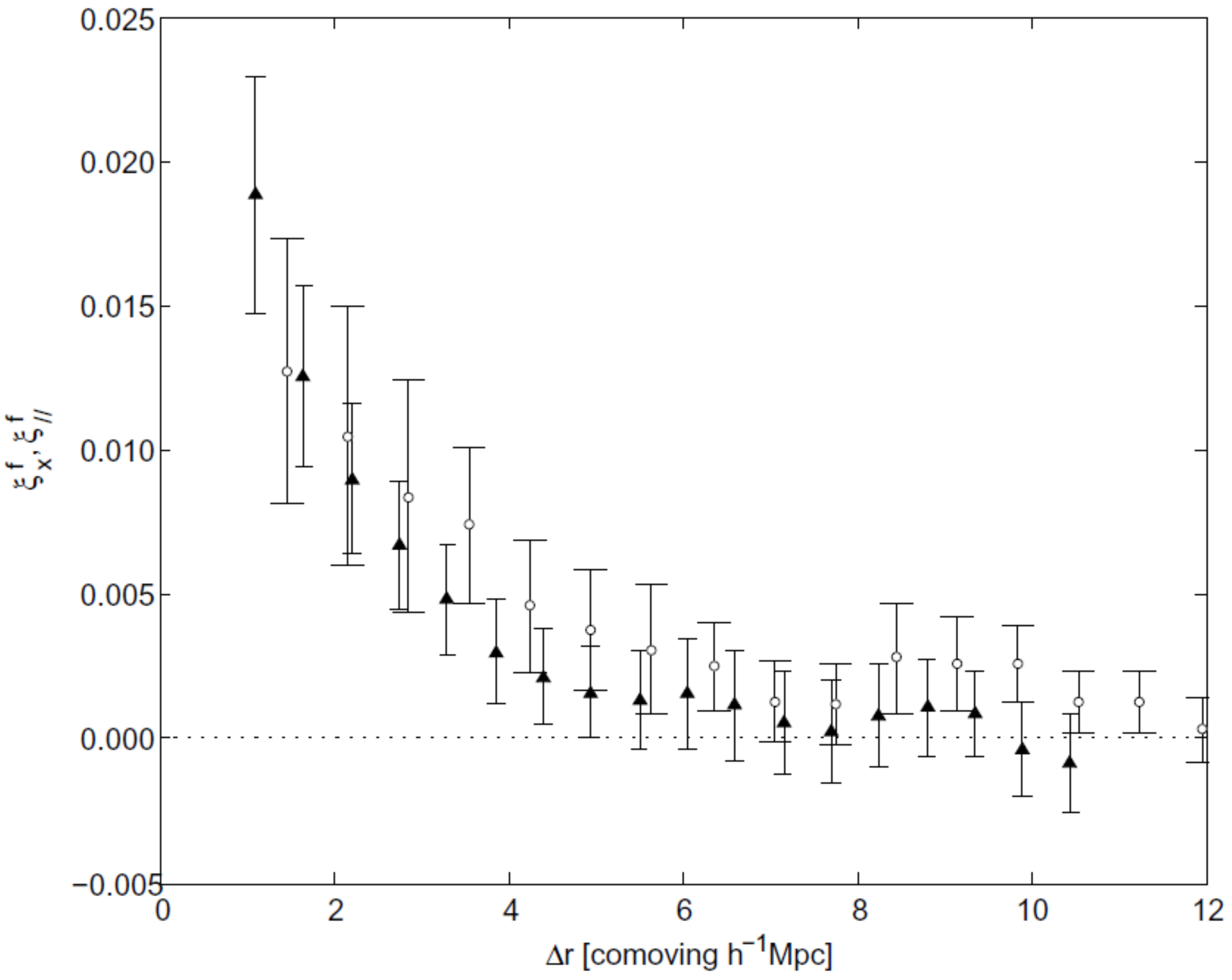}
  \caption{Comparison between Auto and Cross-correlation functions, shown with triangles and
  circles, respectively. Errorbars show the uncertainties related to the simulated 
  samples.}\label{fig:auto-cross} 
\end{figure}

We also compared the Cross CF with the previously computed Auto CF. As can be 
seen in Fig.~\ref{fig:auto-cross} the two data series show a consistency at the  
1~$\sigma$ level, supporting the hypothesis that the peculiar velocity along the \loss\
is negligible.

\subsection{The Cross correlation coefficient}

A measure of the transverse clustering properties of the IGM which is
less affected by peculiar velocities is the set of the flux Cross correlation
coefficients (CCC),

\begin{equation}
\chi^f_{\times}(\Delta\theta)  = \langle \delta_f(\theta, v_{\pa}) 
\delta_f(\theta+\Delta\theta,v_{\pa}) \rangle, 
\label{ccc}
\end{equation}
 
\noindent
where, every pixel along one \los\ is correlated with the one
face-to-face in redshift space along the paired \los\ and the result
is averaged over all the pixels in the common redshift interval.  

Every pair of QSOs at angular separation $\Delta\theta$ gives one
value of $\chi^f_{\times}(\Delta\theta)$, and a sample with several
pairs at different separations, as in our sample, gives an estimate of the
correlation function.  
At a given redshift, the angular separation $\Delta\theta$ corresponds
to a velocity separation $\Delta v_{\perp}=c\,F(z)\,\Delta\theta$,
where $c$ denotes the speed of light, and $F(z)$ is a dimensionless
function of redshift that includes all the dependence on the global
cosmological metric. 
In the cosmological model that we have adopted: 

\begin{equation}
F(z)  =  \frac{E(z) \int_0^z [{\rm d}z'/E(z')]}{(1+z)}. 
\end{equation}

\noindent
where $z$ is the mean redshift of the overlapping \Lya\ region from a pair.
We computed the range of velocity separations covered by each of our pairs of spectra 
then we grouped the pairs in velocity bins of variable width and computed the average 
CCC for every group. Given the small number of pairs 
in each group (a maximum of three QSO pairs) the uncertainties associated with these 
determinations can be only computed by applying a simple error propagation. However, 
these uncertainties would account only for the pixel statistics and the noise associated 
with the transmitted flux, but would be not representative of the true error due to the 
cosmic variance. 
In the case of the simulated spectra, we had 50 realizations of each of our QSO pairs, 
so we could obtain in every velocity interval defined for the observed pairs an average 
CCC with its error, that in this case is the standard deviation of the distribution of values.

\begin{figure}
\includegraphics[height=7truecm,width=8truecm]{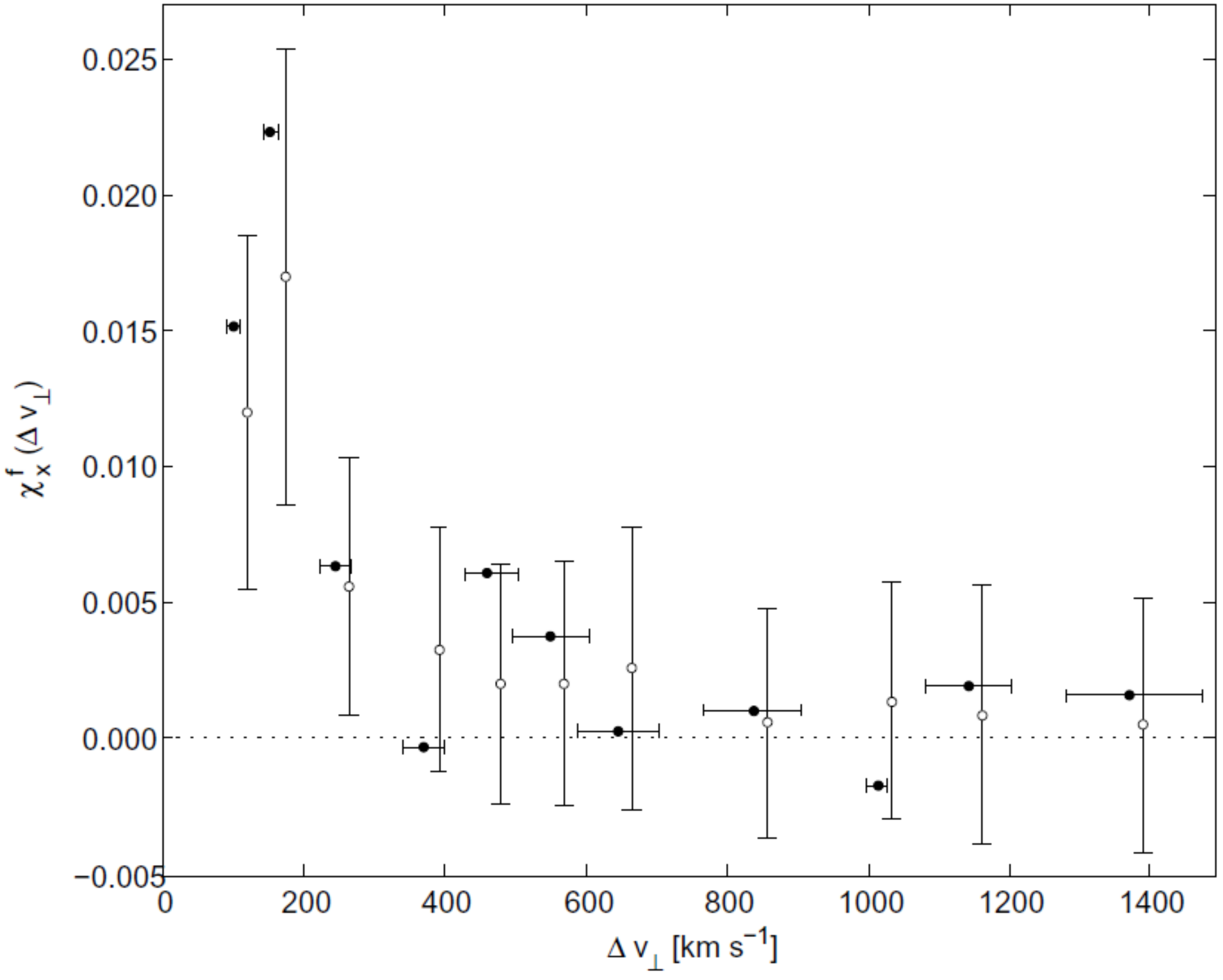}
  \caption{Comparison of the Cross correlation coefficients 
  for our sample of observed spectra (solid dots) and of
  simulated ones (empty circles) as a function of the velocity
  separation, $\Delta v_{\pe} = c\,F(z)\,\Delta\theta$,    
  corresponding to the angular separation, $\Delta\theta$, of the QSO
  pairs. 
  Simulated data are slightly shifted in velocity for clarity.  
  Error bars on the observed values along the x-axis represent 
  the velocity range covered by the considered pairs.}\label{fig:CCC} 
\end{figure}

The results are shown in Fig.~\ref{fig:CCC}. The CCC computed from the observations are 
in excellent agreement, at the $ 1\ \sigma$ confidence level, with the
simulations. 

A comparison can be done with the same results obtained in Paper I: 
thanks to the increased SNR for some of the spectra in 
the present sample, in particular the S1-S3 pair, 
the enhanced clustering signal measured in Paper I 
with the cross correlation coefficient at a transverse 
velocity separation $\Delta v_{\pe} \sim 500$ \kms\ is no longer
significant.

\begin{figure}
\includegraphics[height=6.5truecm,width=8truecm]{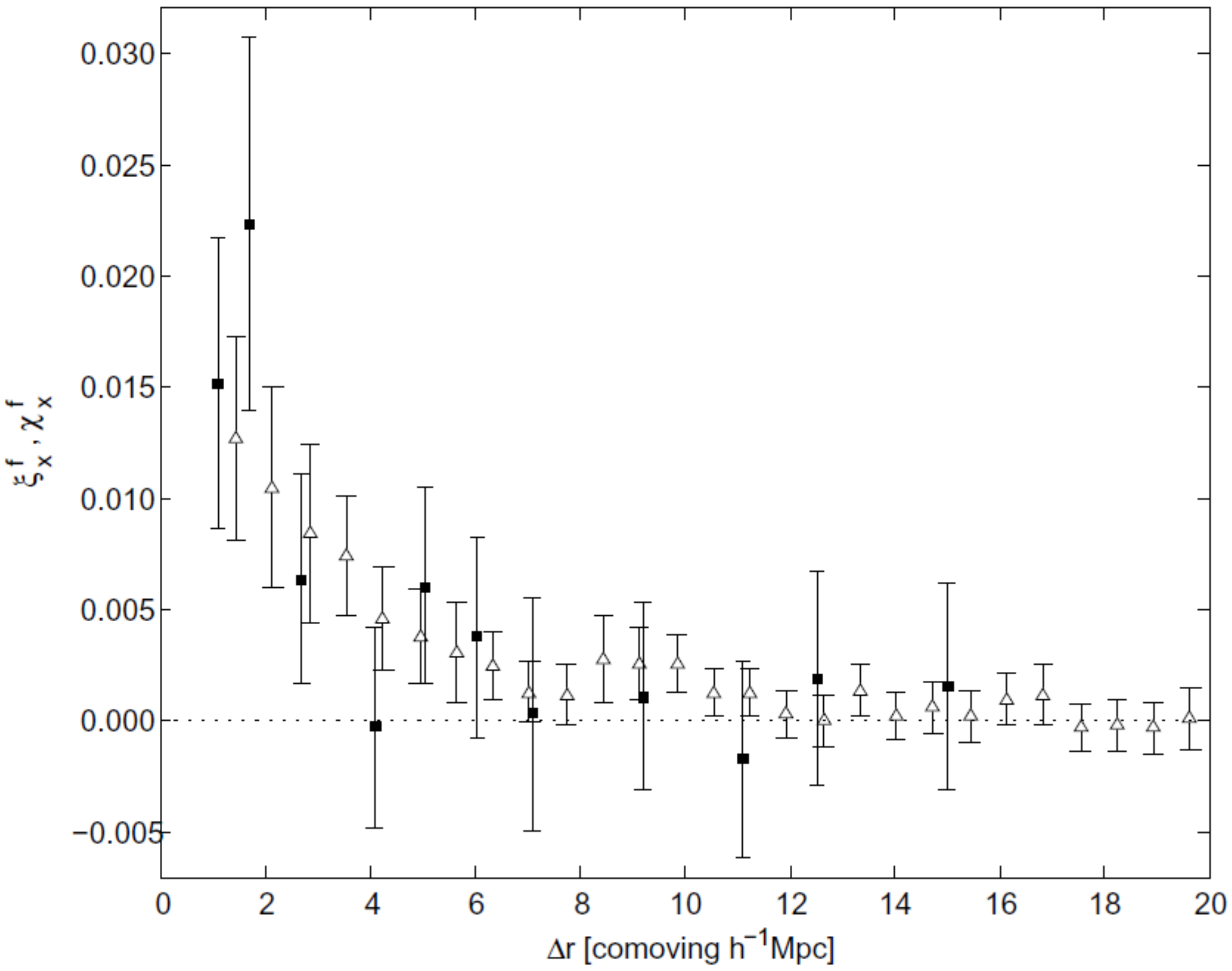}
  \caption{Cross correlation function of the transmitted flux,
    $\xi^f_{\times}$, for our sample of pairs and groups of QSOs
    (empty triangle, slightly shifted in $\Delta r$) compared with the
    observed Cross correlation 
    coefficients, $\chi^f_{\times}$ (squares), as a function of spatial
    separation (see text). Error bars both on $\chi^f_{\times}$ and
    $\xi^f_{\times}$ have been 
    determined from simulations. }\label{fig:CCC-cross} 
\end{figure}

We conclude this section by comparing the CCC with the Cross CF. 
 Both estimators measure the clustering properties of the IGM in
  the transverse direction with respect to the \los. However, the
  Cross CF could be affected by the peculiar velocities in the
  $z$-direction while their effect should be negligible for the CCC.
 
Concerning the CCC, the angular separation $\Delta\theta$ between two QSO \loss\ was 
transformed into a comoving spatial separation, $\Delta r$, with the formula

\begin{equation}
\Delta r = \frac{c \Delta\theta}{H_0} \int_0^z \frac{{\rm d}z'}{E(z')}.
\end{equation}

\noindent
where $E(z)$ was defined above and here $z$ is the mean redshift of the \Lya\ 
interval considered for each pair.
Assigning, at both the CCC and the Cross CF observed data points, the error bars computed 
from the simulations, there is an agreement within $1\ \sigma$ between the two 
series of points (see Fig.~\ref{fig:CCC-cross}), confirming the hypothesis that the 
peculiar velocity along the \loss\ are negligible.
The large variations from one data point to the other 
in the CCC should be due mainly to the small number of QSO pairs (between one and three) 
contributing to each point. On the other hand, the smoothness of $\xi^f_{\times}$ is 
artificially increased by the fact that the values in the different bins are not independent.

All the results reported in this section are in agreement with the same
quantities derived from other larger sample of QSO pairs \citep[see for e.g.][]{rollinde03, coppolani06}.


\section{Coincidences of the absorption lines}

\subsection{The \Lya\ line coincidences}

The Cross CF and the CCC are two powerful tools which allow us to
measure the statistical properties of the IGM distribution in the real
3D comoving space. By definition, they account for the correlation
between pairs of \Lya\ forests in close QSO spectra. What we want
to study now is the simultaneous correlation among three or more
\Lya\ forests. A natural way to extend the Cross CF of the transmitted
flux to more than two \loss\ is the addition of a third
parameter, related to the third QSO at angular separation $\Delta\theta$'
with respect to the first. This operation leads to a function of three
variables, the three reciprocal comoving distances between the three considered
pixels, which is computation time demanding and difficult to interpret.

In the previous section, we have taken advantage of the interpretation
of the \Lya\ forest which allows us to map directly the transmitted flux
along the \los\ into the IGM density field.  
A boost in the signal of the Cross CF between two \loss\ is due to the
presence in redshift space of two aligned, or very close, \Lya\ lines 
belonging to the two considered spectra. 
On this basis, we can provide a measure of the cross correlation between three
\Lya\ forests by searching for triplets of \Lya\ lines, belonging to three different 
spectra, aligned in redshift space within a given velocity window. 

This kind of analysis has been applied to the Triplet and to all the
combinations of 3 QSOs that could be formed with the Sextet. 
The adopted procedure has been the following:

\smallskip
\par\noindent
1. The lists of \HI\ \Lya\ lines compiled for the QSOs in our sample
were considered in the redshift range between the \Lyb\ emission (or
the shortest observed wavelength, when the 
\Lyb\ was not included in the spectrum) and 5000 \kms\ from the
\Lya\ emission (to avoid proximity effect due  
to the QSO).
\par\noindent
2. Each pair of lines with a velocity separation $\Delta v \le 100$ \kms\ has 
been replaced by a single line with central wavelength equal to the average
value of the parent lines weighted on the EW. This velocity threshold
has been chosen on the basis of the characteristic width of 
\Lya\ lines, $\sim 25-30$ km/s \citep[see e.g.][]{kim02}. Furthermore,
this is also the velocity scale corresponding to the Jeans length,
which sets the characteristic dimension of \HI\ absorbers.
\par\noindent
3. Triplets of lines, each one belonging to a different \los, have been considered 
and the velocity difference between the largest and smallest redshift has 
been computed. 
This operation has been done for the \Lya\ lines in the three
\loss\ of the Triplet and in all the triplets of  
\loss\ (20 possible combinations) provided by the Sextet.  
Then, all the measures of velocity difference lower than 1000 \kms\ have been divided 
into velocity bins of 100 \kms\ and the related histogram with the number of occurrences 
for each bin has been computed. 
\par\noindent
4. Next, the previous three steps have been repeated for a sample of $10^3$
mock lists of lines built in the following way. 
In order to take into account the varying number density of detectable
lines along the \Lya\ forests, due to the varying SNR, each forest has
been simulated in chunks of about 200 \AA. 
In each mock chunk, the number of simulated lines has been determined from a Poissonian distribution 
centred on the number of observed lines in that chunk, while the positions of the mock 
lines have been randomly generated following a uniform distribution within the related 
wavelength range of each chunk. The redshift intervals masked in the observed spectra 
were masked also in the simulated ones. The EWs of the mock lines have been randomly 
chosen among all the EWs measured by the fit of the lines in the observed spectra.  
In this way it has been possible, for each velocity bin, to compute the mean and the 
standard deviation of the number of occurrences for synthetic lists of lines.
\par\noindent
5. Finally, we have defined the three point probability excess (PE3) as a function of the 
velocity difference, $\Delta\,v$, according to the following formula: 

\begin{equation}
PE(v)=N_{obs}(\Delta\,v) / N_{sim}(\Delta\,v) - 1
\label{eq:PE}
\end{equation}
   
\par\noindent

\begin{figure}
\includegraphics[height=7truecm,width=8truecm]{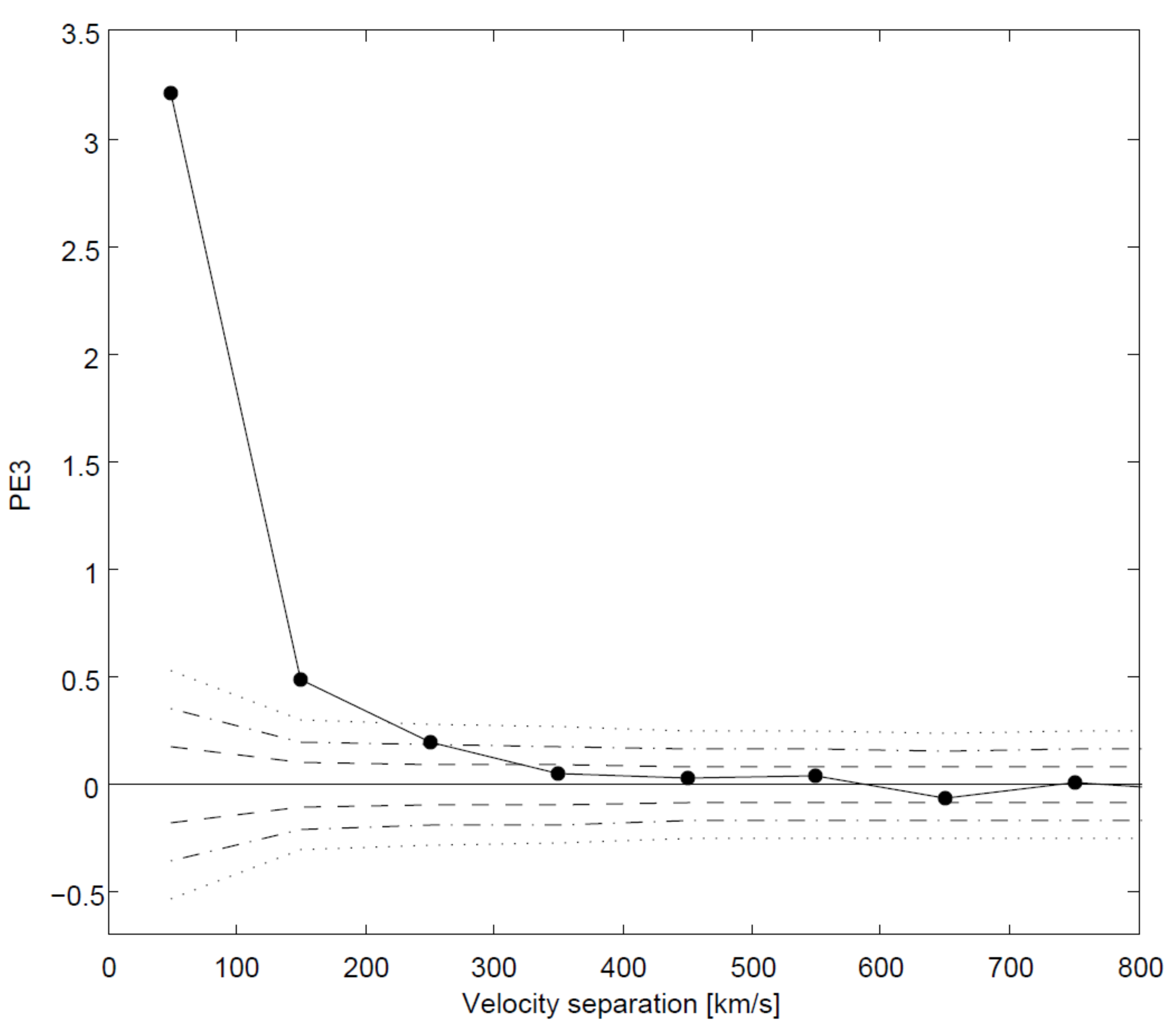}
  \caption{Three point probability excess (PE3) as a function of the 
  velocity difference of the three \Lya\ lines (dots).
  Values referring to a significancy level of 1, 2 and 3~$\sigma$ 
  levels are reported too (dashed, dot-dashed and dot lines respectively).}
  \label{fig:PE3} 
\end{figure}

The resulting PE3 is reported in Fig.~\ref{fig:PE3}, together with the
1, 2 and 3~$\sigma$ confidence levels. The PE3 is non-zero at 
a 2~$\sigma$ level up to a velocity difference of $\sim 250$ \kms. 
Most of the signal of the PE3 is due to the large number (26) of 
coincidences produced by the S3-S5-S6 QSOs triplet which is also the 
closest triplet (mean angular separation of 2.02 arcmin corresponding
to $\sim 2$ $h^{-1}$ comoving Mpc).

\begin{figure}
\includegraphics[height=7truecm,width=8truecm]{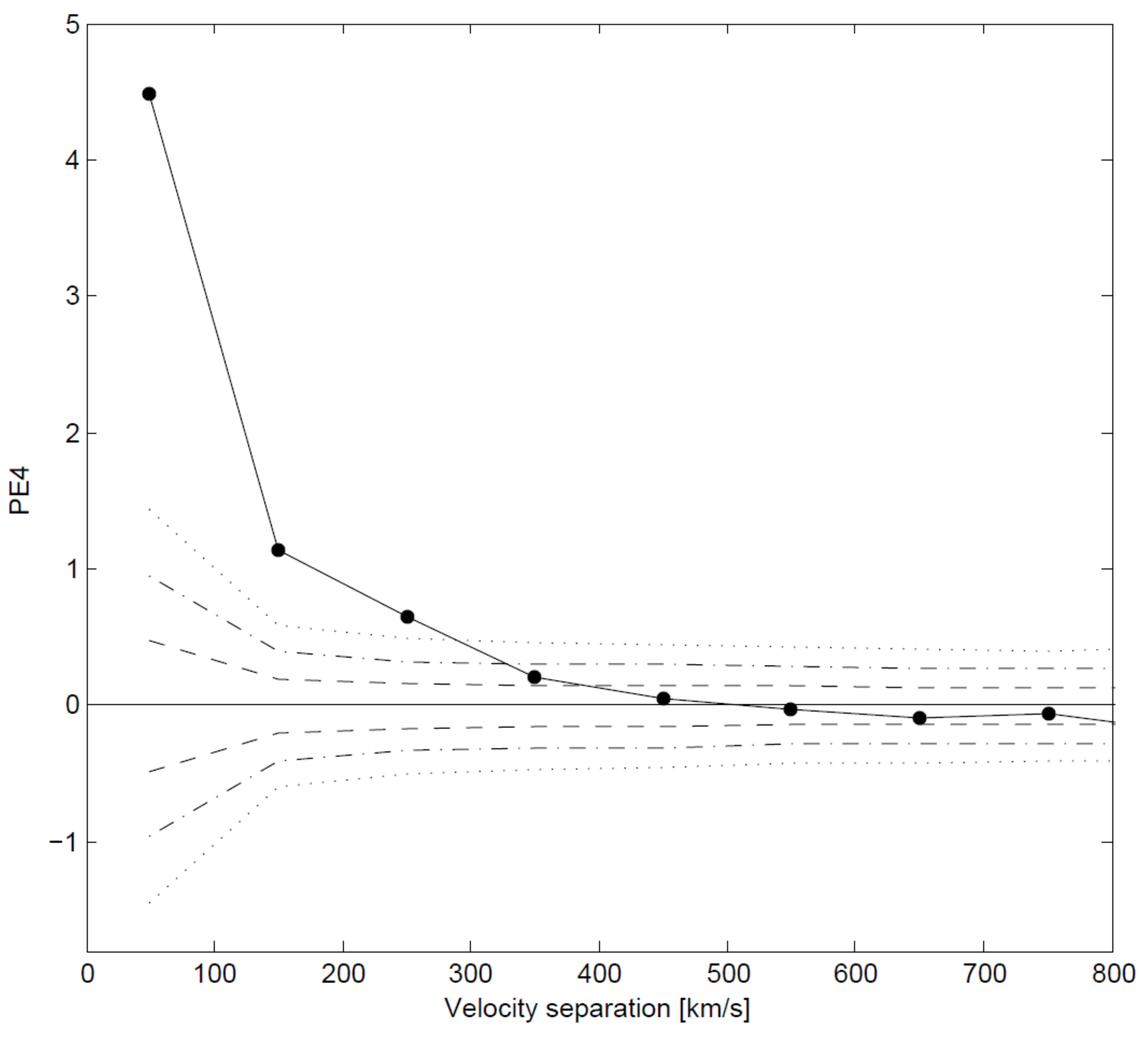}
  \caption{Four point probability excess (PE4) as a function of the 
  velocity difference of the three \Lya\ lines (dots).
  Values referring to a significancy level of 1, 2 and 3~$\sigma$ 
  levels are reported too (dashed, dot-dashed and dot lines respectively).}
  \label{fig:PE4} 
\end{figure}

\begin{figure}
\includegraphics[height=12truecm,width=8truecm]{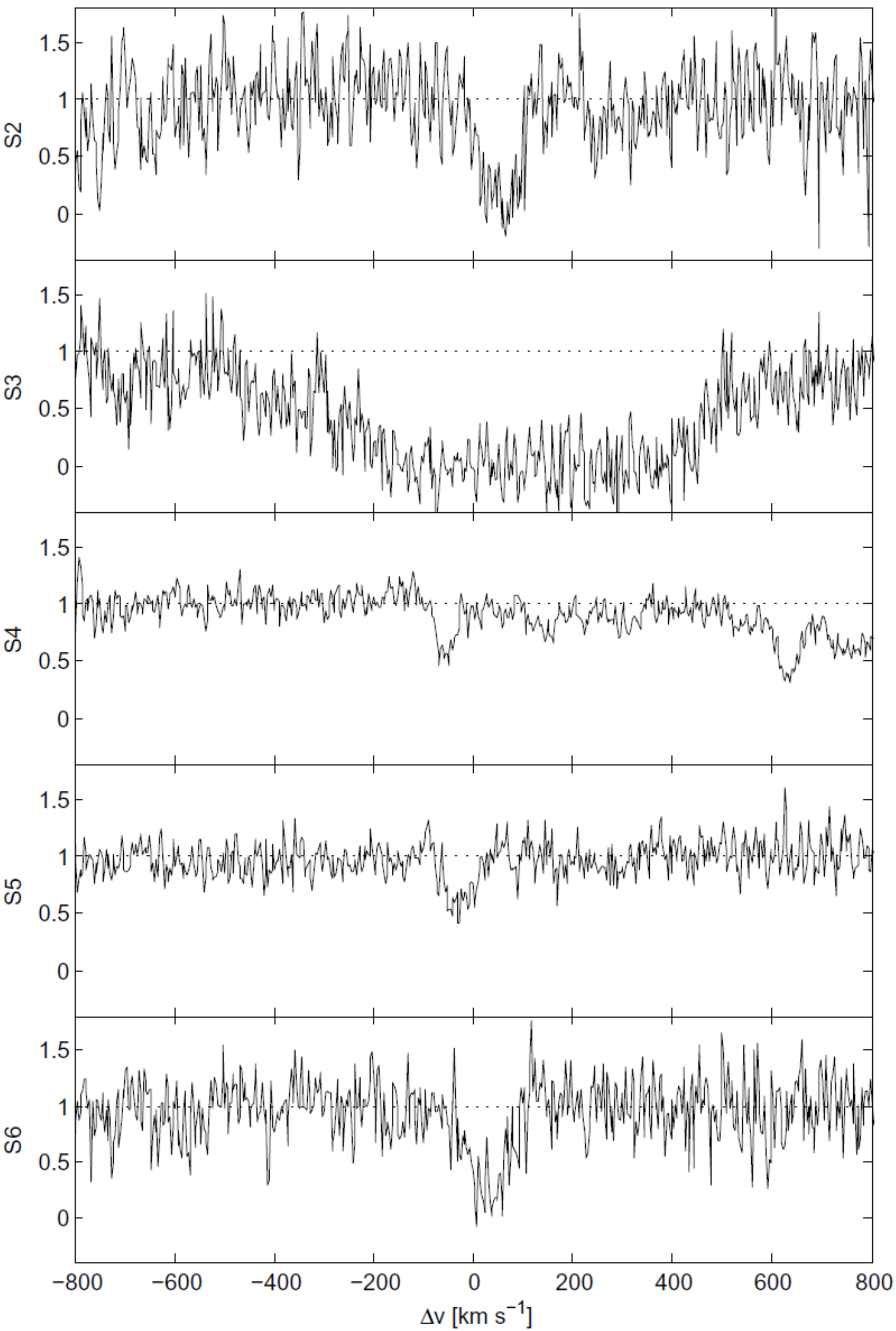}
  \caption{Portions of the spectra of the QSOs S2, S3, S4, S5 and S6 
  centred at 3435 \AA\ showing the aligned \Lya\ lines.}
  \label{fig:Filament} 
\end{figure}

Fig.~\ref{fig:PE4} shows the probability excess considering quadruplets of 
\Lya\ lines. A significant signal at more than 3~$\sigma$ level
is measured up to a velocity difference of $\sim 250$ \kms.
Besides, one group of five coincident lines within 100 \kms\ in the 
S2-S3-S4-S5-S6 QSOs is observed at a mean redshift of 1.825, 
an occurrence that has a probability P=0.013 to arise from a random 
distribution of lines. The portion of spectra where these five lines 
fall are reported in Fig.~\ref{fig:Filament}.

In particular, it is possible to observe in the spectrum of the S3 QSO 
the presence of a Damped \Lya\ system (DLA): the fitted Voigt profile gives a 
column density value of $\log N$ = 20.6. This DLA is associated with several 
metallic ion absorption lines found in the redder part of the spectrum. 
Indeed at the same redshift we have found evidence of \CIV, \FeII, \SiIV,
\SiIII, \SiII, \AlIII\ and \AlII. 
This correlated \HI\ absorption across \loss\ separated by $\sim 14$ 
$h^{-1}$ comoving Mpc could be interpreted as due to a gas filament, but 
other explanations are possible such as detected \HI\ halos of clustered 
galaxies.

\subsection{The \CIV\ lines coincidences}

\begin{figure*}
\includegraphics[height=7truecm,width=8.5truecm]{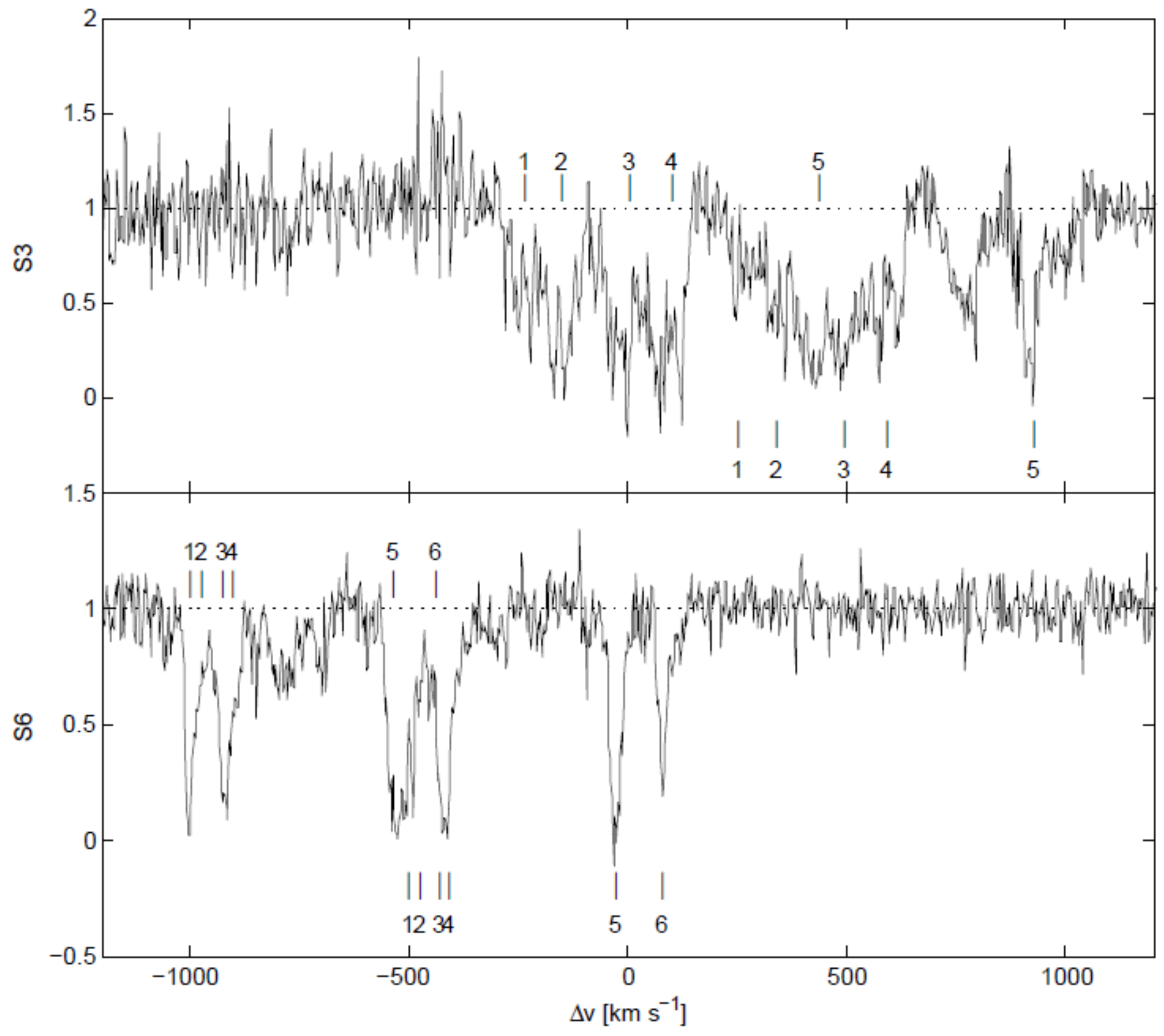}
\includegraphics[height=7truecm,width=8.5truecm]{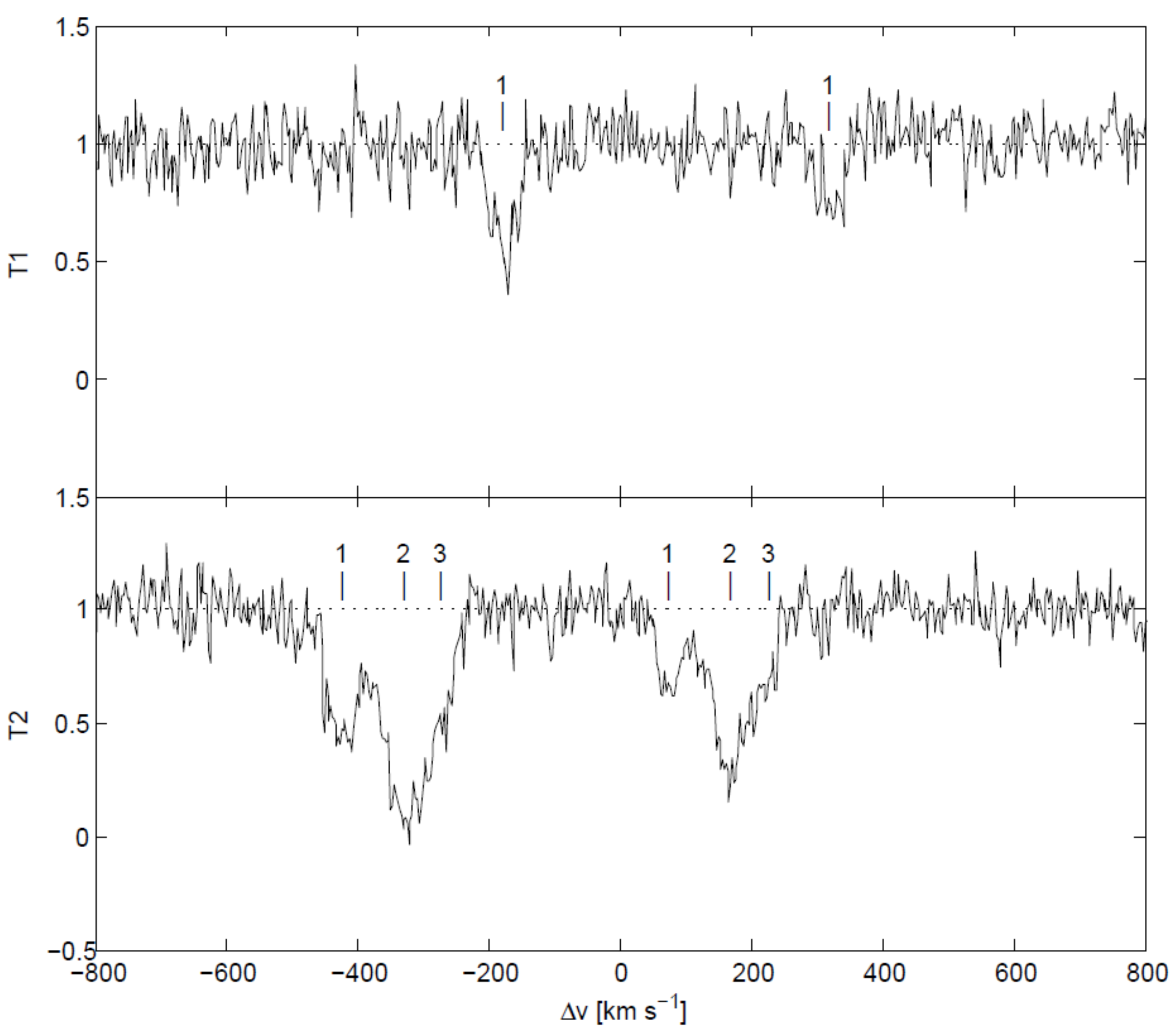}
  \caption{Portions of spectra representing the \CIV\ systems found to 
  be in coincidence between the S3 and S6 QSOs (left plots) and the 
  T1 and T2 QSOs (right plots). The single \CIV\ components are marked 
  with increasing number. The same number is related
  to both the \CIV\ lines.}\label{fig:CIV_coinc} 
\end{figure*}

We have also looked for correlation across two \loss\ for \CIV\ absorbers. 
Given the characteristics of our sample, which allow us to investigate comoving 
scales larger than $\sim 1 h^{-1}$ comoving  Mpc, we do not expect \CIV\ 
coincidences due to the same \CIV\ cloud, whose typical size should be of 
the order of tens of kpc \citep{rauch01}. Our \loss\ could pierce \CIV\ gas 
connected with different galaxies in groups or clusters \citep{francis01}.

The procedure to identify \CIV\ coincidences is similar but simpler 
than that previously employed for the \Lya\ lines. 
The first two steps are the same, with the difference that here 
lines closer than 500 \kms\ have been merged together. 
Then, we searched for coincidences of \CIV\ merged absorbers lying along the 
redshift space in windows smaller than 500 \kms.
According to \citet{scannapieco06} this velocity scale
corresponds to the characteristic size of coherent \CIV\ structures.

In order to establish the significance of our observation, we have applied our 
algorithm for coincidences to $10^4$ groups of synthetic list of lines.
The method for creating the synthetic list of lines is the same as adopted 
before in the case of the mock list of \Lya\ lines.
With the total set of simulated results, we could find the number 
$N(n \ge n_{0})$ of cases in which, in a random distribution of \CIV\ absorbers 
along the \loss, the number of chance coincidences $n$ is equal to or greater 
than the number of coincidences observed in our sample, $n_{0}$. 
Then the related probability $P$ was simply computed using the ratio 
between $N(n \ge n_{0})$ and the total number of simulations, that is $10^4$.
Finally, we computed the CL of our observation simply with the formula  
$CL = 1 - P(n \ge n_{0})$.

We have found only two coincidences, in the T1-T2 and S3-S6 pairs, at a redshift of 
2.062 and 1.486 respectively and separated by a comoving distance of 4.4 and 2.8 h$^{-1}$ comoving Mpc.
The statistical significance of these detections are 90\% and 85\% respectively. 
Coincidences have also been found between the T2 QSO redshift and a T3 absorber 
(with a separation of 9.6 h$^{-1}$ comoving Mpc) and between the S6 QSO redshift and a S3 
absorber (with a separation of 2.8 h$^{-1}$ comoving Mpc).
These coincidences are in agreement with the fact that QSOs reside in over-dense regions of 
the Universe \citep[e.g.,][]{vale08}.
No coincidences among absorbers have been found considering three or more \loss.
The portions of spectra showing the two coincidences are reported in 
Fig.~\ref{fig:CIV_coinc}.


\section{Common under-dense regions}

The two groups of QSOs in our sample are well suited to search for 
large under-dense regions in the IGM.  
Previous attempts to detect under-dense regions in the \Lya\ forest
have searched for regions without absorption lines along a single
\los. They looked for portions of spectra where the observed number of
absorption lines was significantly smaller then the expected number
from synthetic spectra \citep{carswell87,crotts87,ostriker88,cristiani95,kim01}.  

\begin{figure*}
\includegraphics[height=8truecm,width=8.5truecm]{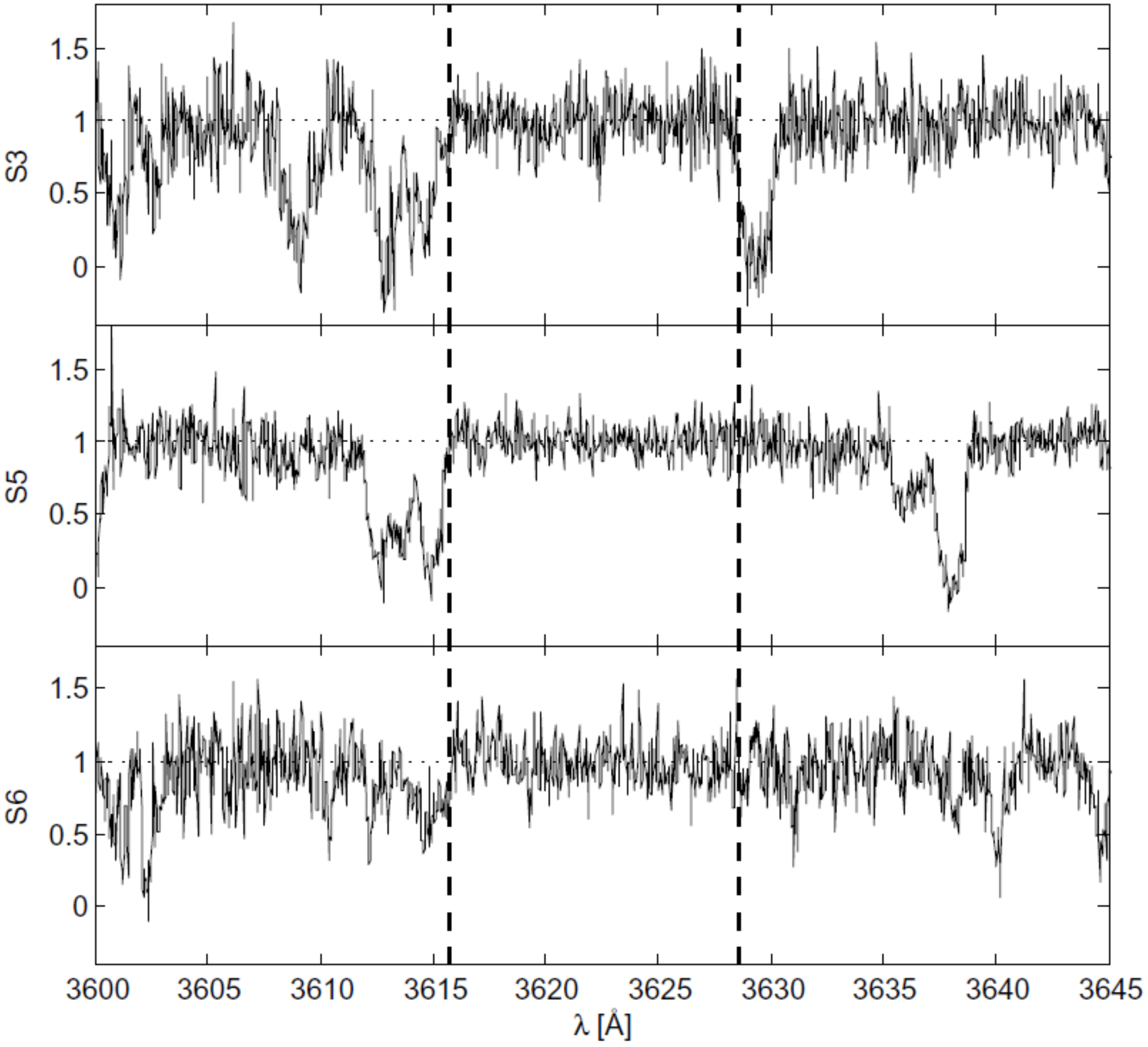}
\includegraphics[height=8truecm,width=8.5truecm]{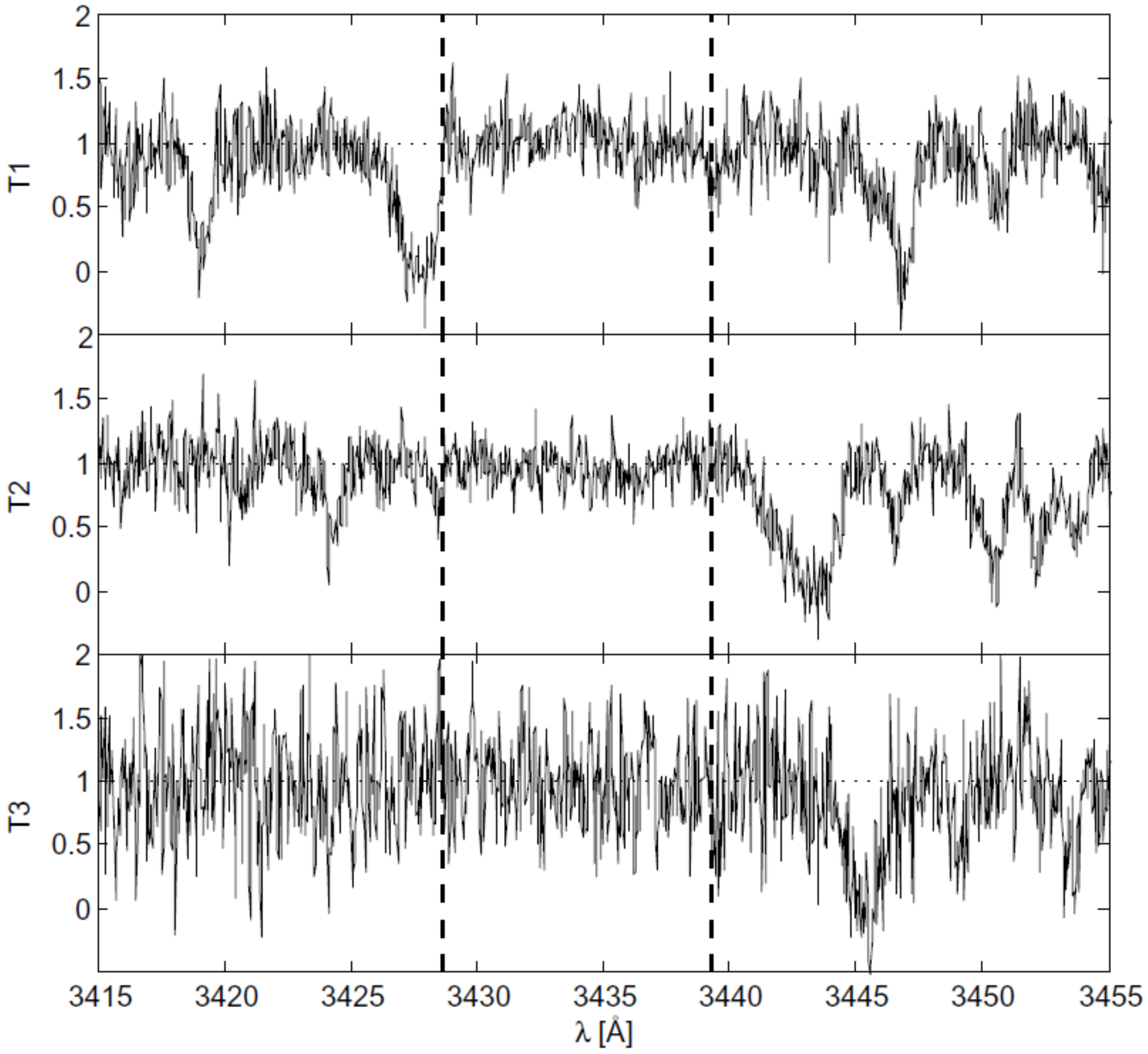}
  \caption{Spectra of the two close groups of quasars (S3-S5-S6 on 
  the left and the Triplet on the right) centred on the common under-dense 
  regions. The extension of the common under-dense regions are shown with dashed vertical 
  lines.}\label{fig:voids} 
\end{figure*}

 A strong improvement of the statistical significance of each detection
can be provided by the simultaneous presence of an under-dense region 
along several close \loss. This would allow the detection of smaller 
under-dense regions maintaining a high confidence level. 
As done in the previous section, we focused our attention on searching 
for common under-dense regions among the \loss\ of the Triplet and among 
all the possible groups of \loss\ provided by the Sextet.
 
Following \citet{rollinde03}, we can define an under-dense region along a 
single \los\ as a portion of the spectrum between $\lambda_{i}$ and $\lambda_{e}$, 
where the normalized flux ${\it f}$ is higher than
the specific threshold $(\bar{f} - \sigma_f)$, where $\bar{f}$ is the
average flux in the \Lya\ forest and $\sigma_f$ is the value of the flux uncertainty. 
Note also that at $z\sim 2$ regions above the mean density correspond on average to
regions below the mean flux level of high resolution spectra \citep{viel2008}. 
Since the SNR of the spectra analyzed by
\citet{rollinde03} is much larger than that of our spectra ($15-70$ vs.
$4-15$ respectively), this simple procedure for the detection of
under-dense regions cannot be adopted in the present case. Indeed,
the fluctuations due to the flux noise cause the splitting of single
large under-dense regions into several smaller regions, preventing
their detection.

\begin{figure}
\includegraphics[height=7truecm,width=8truecm]{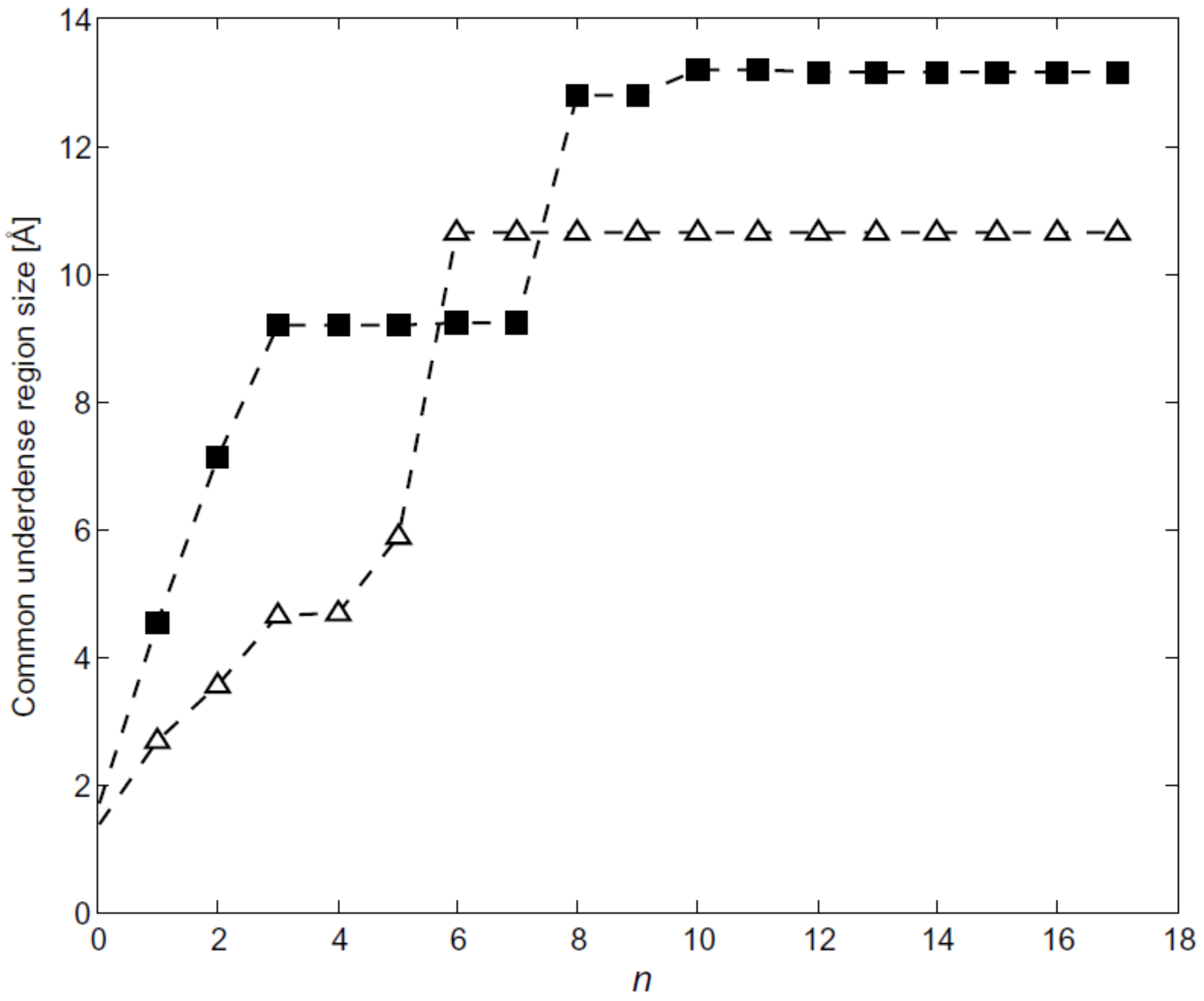}
  \caption{Size of the common under-dense regions of Triplet group 
  (empty triangle) and the S3-S5-S6 group (filled square) as a function of 
  the number $n$ of times the spectra have been smoothed.}\label{fig:smooth} 
\end{figure}

In order to reveal the presence of under-dense regions in our
spectra, we have developed the following method. The spectra are smoothed
with a median filter characterized by a very short kernel length, only
three pixels. After the smoothing, common under-dense regions are looked
for with the method described above (always adopting the same average
flux and flux uncertainty). Then, the spectra are smoothed 
again and again searched for common voids. This process is iterated
several times.  The results are shown in
Fig.~\ref{fig:smooth} where we have reported the size of the bigger
common under-dense region found in the Triplet and in the S3-S5-S6 group
of Sextet QSOs as a function of the number of times the spectra have been smoothed.
We report only the result concerning the S3-S5-S6 group because it is the 
only one in which a significant under-dense region has been detected.

We can see how, after an initial linear growth of the size of the common
under-dense region, a step behaviour occurs in both the series of data. 
Steps occur whenever two separate under-dense regions are merged together, 
that is when the absorption line (both real or spurious) lying between them 
is smoothed out by the n-th action of the median filter. 
We can notice also how, after $\sim 10$ smoothings, the size of the 
under-dense regions in both the Triplet and the S3-S5-S6 group reaches
an ''asymptotic'' value. Each asymptotic rate corresponds to 
the size of a common under-dense region enclosed between two strong 
lines (not necessarily belonging to the same spectrum) requiring a
large number of smoothings to be cancelled. 

A series of simulations has been carried out in order to evaluate the minimum value 
of the column density $N($\HI$)$ of the \Lya\ lines surviving after 6 and 8 smoothing 
processes in the Triplet and the S3-S5-S6 group, respectively.
First, a mock spectrum has been created placing several lines characterized by an 
increasing $\log N($\HI$)$, from 12 to 15 by steps of 0.05, and central wavelengths
separated by 10 \AA. 
Line profiles have been modeled with a Gaussian, since we were interested only in 
the centre of the lines and not in reproducing correctly their wings. 
The Doppler parameter $b$ was set to a value of 20 \kms \citep{kim01}; 
the dependence of the simulation results from this parameter is negligible. 

\begin{figure}
\includegraphics[height=7truecm,width=8truecm]{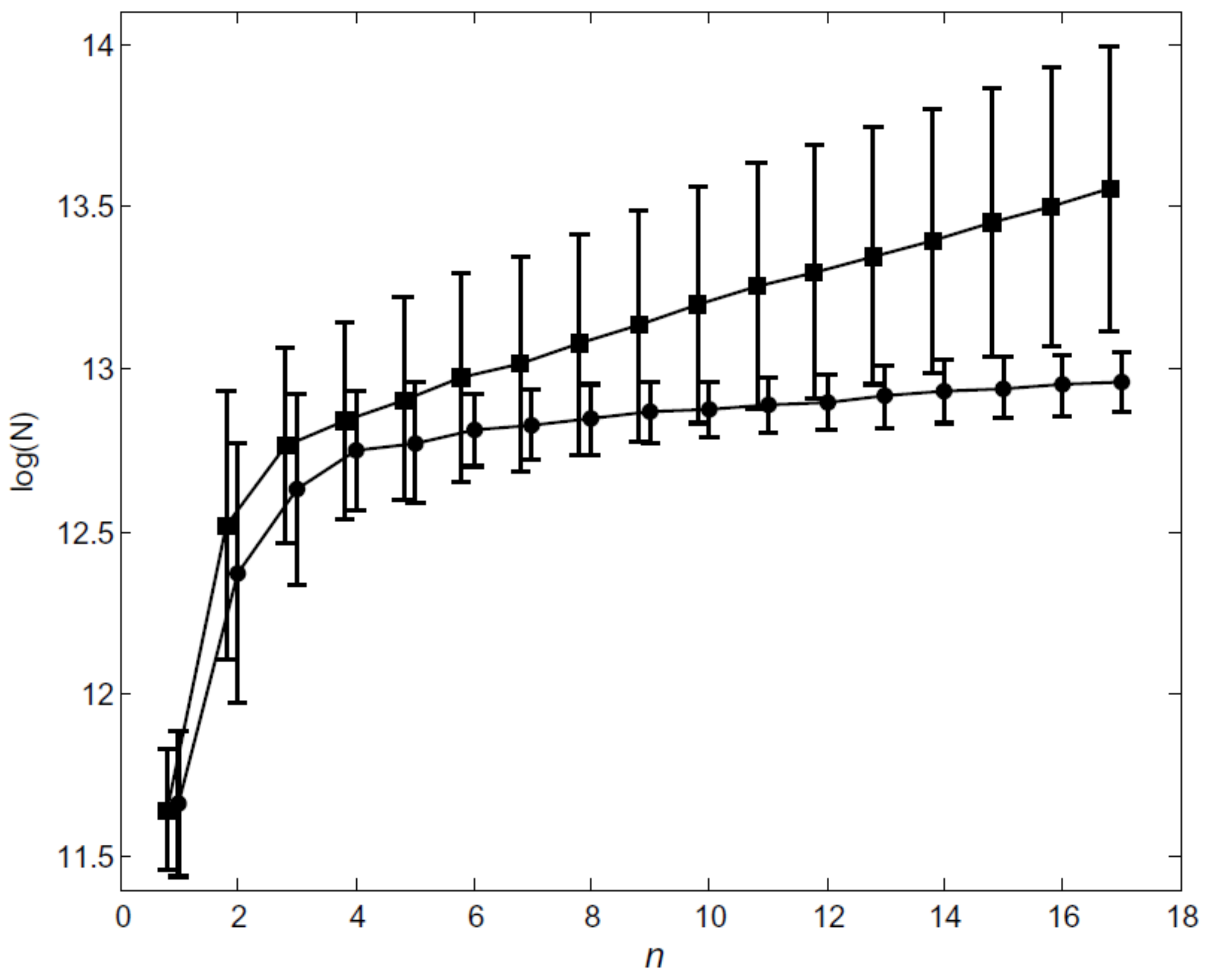}
  \caption{Minimum values of the \HI\ column density of the \Lya\ lines left after
   a given number $n$ of smoothing with a three pixel long filter (circle) and after one 
   smoothing with a (2$n$+1) pixel wide filter (square, 
   slightly shifted along the $x$-axis for clarity). }
  \label{fig:N-Smooth} 
\end{figure}

White noise has been added to the spectrum in order to obtain $SNR=7$, 
roughly consistent with the observed SNR values of the Triplet, and of
the S3, S5 and S6 spectra in the segments where the common under-dense regions are situated.
Then, the length of the void has been measured, namely the portion of mock spectrum where 
${\it f} \ge (\bar{f} - \sigma_f)$, after each smoothing. Since the lines are ordered by 
increasing column density and separated by 10 \AA, each increase in 
void length of $\sim 10$ \AA\ corresponds to the vanishing of a specific line with known
value of $N($\HI$)$.
Repeating this procedure $10^3$ times, we have computed the values, and the related uncertainties, 
of the column density of the lines washed away after a given number of smoothings.
The effect of having different SNR in the QSOs spectra (Fig.~\ref{fig:voids}) instead of
a fixed value of $SNR=7$ is smaller than the error bars reported in Fig.~\ref{fig:N-Smooth}.

The same kind of simulations have been used to compare the action of the short scale 
filter iterated $n$ times with respect to that of a median filter of length (2$n$+1) 
pixel employed once. Both the series of data are shown in Fig.~\ref{fig:N-Smooth}, with 
the related error bar, as a function of $n$. The trend followed by the two series at 
large $n$ is different: for the short scale filter iterated $n$ times the smoothing 
effect is relatively lower and the growth of the column density of the lines smoothed 
away is flatter. Furthermore, the related uncertainties are much smaller. This behaviour 
helps us to estimate more accurately the lower column density of the surviving lines, because 
the range of $\log N($\HI$)$ between 12.5 and 13 is sampled more accurately with respect 
to the upper curve.

Fig.~\ref{fig:N-Smooth} (filled circles) shows that the asymptotic 
size of the voids observed in the Triplet and in the S3-S5-S6 group 
are referred to spectra where the absorption lines characterized 
by $\log N($\HI$)\la 12.8$ have been smoothed away.

The two largest under-dense regions common to the S3, S5 and S6 \loss\ and to the 
Triplet \loss\ are 12.8 \AA\ (corresponding to 10.7 $h^{-1}$ comoving Mpc) and 10.6 \AA\ 
(or 8.8 $h^{-1}$ comoving Mpc) in size, respectively. They are shown in Fig.~\ref{fig:voids}.
The sizes of the related under-dense regions along the single lines of sight 
are: 13.4, 19.8 and 15.2 \AA\ for the S3, S5 and S6 \loss; 10.6, 12.4 and 14.2 \AA\ 
for the \loss\ of the Triplet. As already said, no common under-dense regions have been 
found considering any other combination of \loss\ in the Sextet.

Assuming that the under-dense regions have a spherical geometry
\citep[again following][]{rollinde03}, a void is then  
defined as the largest sphere that may be included
inside one connected region. We searched for the largest
spherical region included in the three individual under-dense regions. 
This is equivalent to the requirement that the surface of the under-dense
region matches the six edges of the three segments. We allow an uncertainty 
of few \AA\ in the position of the surface along each \los. This takes into
account the fact that observations are in redshift space, so that the border
of the under-dense region may have a peculiar velocity that shifts
its position in wavelength, and that a real under-dense region  
probably does not have a simple boundary surface.
This set of equations can constrain a sphere that has four degrees of 
freedom. 
Inside the sphere, the IGM will be under-dense 
along each \los\ and on its surface it will be over-dense at six
positions identifying the edges of the individual under-dense regions.
When we applied this procedure to the S3-S5-S6 group, the under-dense 
region can be parametrized with a sphere of 6.75 $h^{-1}$ comoving Mpc radius.
On the other hand, due to the positions of the members of  Triplet in the 
sky, a spherical parametrisation of the void is not possible in this case.

 In order to assign a significance level to the detection of 
these two under-dense regions, we performed $10^3$ simulations of mock spectra
reproducing the two triplets of QSO spectra. First, the lists of central 
wavelengths were created following the same procedure as defined in
Section.~4.1. Then, the line profiles were created
with the observed column density and Doppler parameter distribution
\citep[e.g.,][]{kim02}. Random white noise was added, according to the 
observed SNR of each spectrum in each chunk, and then the spectra were 
smoothed 8 times by the median filter.
We verified that the results of the simulations do not change significantly
for a number of iterations larger than six.
Finally, we computed the distribution function of the probability of having 
one common under-dense region of given dimension along three \loss.
Computing the random probability of having a common under-dense regions
of size equal to or greater than ours, we obtained that the final 
significance levels of our detections are 91\% and 86\% for the 
Triplet and for the S3-S5-S6 group respectively.
These two values are smaller than the significance levels of the common 
under-dense region detected by \citet{rollinde03}. We expect that this
is due to  the fact that we are dealing with only three \loss\ and that our
spectra have a lower SNR.


\section{Conclusions} 

In this paper, we exploited the capabilities 
of high resolution UVES spectra of QSO pairs to study the
3-dimensional distribution properties of baryonic matter in the IGM as
traced by the transmitted flux in the QSO \Lya\ forests. 
Our sample is formed by 21 QSO pairs evenly distributed between
angular separations of $\sim 1$ and 14 arcmin, with \Lya\ forests at
a median redshift $z \simeq 1.8$. 
By calculating the correlation functions we compared the observed 
sample with a set of mock spectra drawn 
from a cosmological hydro-simulation run in a box of $120\ h^{-1}$
comoving Mpc, adopting the cosmological parameters of the concordance
model. The simulated sample reproduces 50 different realizations of the
observed sample (see Section 2 for details). 
Furthermore, particular emphasis has been given to the search for alignments 
and other particular features of the \Lya\ forests in the two QSO groups present in our sample.

In the following, we summarize our main results:

\noindent{\em Two-point statistics}
\begin{enumerate}
\item{The computed correlation functions are in substantial agreement 
with those obtained in our previous paper \citep[][Paper I]{vale06}. 
There is consistency between the clustering properties of matter in 
the IGM calculated in the direction parallel and transverse to the line of
sight using the parameters of the concordance cosmology to map the
angular distance into velocity separation. This is also due to the
relatively large error bars of the computed quantities. As an
implication, peculiar velocities in the absorbing gas are likely to be
smaller than $\sim 100$ \kms.  Matter in the IGM is clustered on
scales smaller than $\sim 300$ \kms\ or about $4\ h^{-1}$ comoving
Mpc. The simulated correlation functions are consistent with the
observed analogous quantities at the $1-2\,\sigma$ level for this 
particular sample.  
}
\item{Thanks to the increased SNR for some of the spectra in 
our sample the enhanced clustering signal measured in Paper I 
with the cross correlation coefficient at a transverse 
velocity separation $\Delta v_{\pe} \sim 500$ \kms\ is no longer
significant.}
\smallskip
\par\noindent
{\em Three or more point statistics}
\item{Significant coincidences of \Lya\ absorptions have been detected among 
the \loss\ forming the Sextet implying the presence of coherent gas structures 
extending $\sim14 h^{-1}$ comoving Mpc. In particular an excess of triplets and 
quadruplets of lines within $\Delta v = 100$
\kms\ has been measured at a significance of 16 and 9 $\sigma$.
Besides, one group of five coincident lines in the S2-S3-S4-S5-S6 QSOs is observed,
an occurrence that has a probability P=0.013 to arise from a random distribution
of lines.}
\item{A method for the detection of under-dense regions in relatively 
low SNR spectra has been developed. One cosmic common under-dense region has 
been detected in each QSO group; in the Sextet the under-dense region has a dimension 
of 10.7 $h^{-1}$ comoving Mpc and a significance level of 91\%, while in the Triplet 
it has a dimension of 8.8 $h^{-1}$ comoving Mpc and a significance level of 86\%.
These values are significantly smaller than those typical of under-dense 
regions detected along single lines of sight. The under-dense region common to the 
lines of sight of the Sextet can be  parametrized by a sphere of radius $6.75\ h^{-1}$ 
comoving Mpc.  
}
\end{enumerate}

This study of the cosmic web environment at $z\sim 2$ will soon
be extended with many more QSO spectra either at medium resolution
with the X-Shooter spectrograph \citep{kaper09} or/and 
with the $R\sim 10^5$ QSO spectra
provided by SDSS-III \citep{schlege07}. The ultimate goal is the
characterization of the topology of the IGM in real space in a variety
of environments by using \Lya\ and metal lines. 
On this basis, both  hydrodynamical simulations of structure formation and 
high-resolution spectroscopic samples, like the one presented here, will
provide the link between observational quantities and the underlying
density field and shed light on the impact of systematic effects.

\section*{Acknowledgments}
MC is grateful to Prof. R. Rui for the stimulating discussion on the
right statistical approach and interpretation to be adopted for this
work. Simulations were performed at the COSMOS UK National Cosmology
Supercomputer, sponsored by SGI, Intel and HEFCE.
This research has been partially supported by ASI Contract No.
I/016/07/0 COFIS, INFN PD51 grant and PRIN MIUR "Diffuse baryons in 
the Universe".


\label{lastpage}


\begin{thebibliography}{99}


\bibitem[\protect\citeauthoryear{Aracil et al.}{2002}]{aracil02}
Aracil B., Petitjean P., Smette A., Surdej J., Mucket J. P. and Cristiani S., 2002, A\&A, 391, 1

\bibitem[\protect\citeauthoryear{Ballester et al.}{2000}]{ballester00}
Ballester P., Modigliani A., Boitquin O., Cristiani S.,
Hanuschik R., Kaufer A., Wolf S., 2000, ESO The Messenger,
101, 31

\bibitem[\protect\citeauthoryear{Becker, Sargent \& Rauch}{Becker et al.}{2004}]{becker04}  
Becker G. D., Sargent W. L. W., Rauch M., 2004, ApJ, 613, 61

\bibitem[\protect\citeauthoryear{Bianchi et al.}{2003}]{bianchi03}  
Bianchi S., Cristiani S., Kim T. S., D'Odorico S., 2003, Astronomy, Cosmology and Fundamental Physics, 4-7 March 2002, p. 417

\bibitem[\protect\citeauthoryear{Brian \& Machacek}{2000}]{brianmachacek00}
Bryan G. L., Machacek M. E., 2000, ApJ, 534, 57

\bibitem[\protect\citeauthoryear{Carswell \& Rees}{1987}]{carswell87}
Carswell R. F., Rees M. J., 1987, MNRAS, 224, 13

\bibitem[\protect\citeauthoryear{Coppolani et al.}{2006}]{coppolani06}
Coppolani F., Petitjean P., Stoehr F., Rollinde E.,Pichon C., Colombi S.,
Haehnelt M. G.,Carswell B. and Teyssier R., 2006, MNRAS, 370, 1804


\bibitem[\protect\citeauthoryear{Cristiani et al.}{1995}]{cristiani95}
Cristiani S., D'Odorico S., Fontana A., Giallongo E., Savaglio S., 1995, MNRAS, 273, 1016

\bibitem[\protect\citeauthoryear{Croft et al.}{2002}]{croft02}
Croft R. A. C., Weinberg D. H., Bolte M., Burles S., Hernquist L., 
Katz N., Kirkman D., Tytler D., 2002, ApJ, 581, 20

\bibitem[\protect\citeauthoryear{Crotts}{1987}]{crotts87}
Crotts A. P. S., 1987, MNRAS, 228, 41

\bibitem[\protect\citeauthoryear{Dav\`e et al.}{1999}]{dave99}
Dav\`e R., Hernquist L., Katz N., Weinberg D. H., 1999, ApJ, 511, 521

\bibitem[\protect\citeauthoryear{D'Odorico et al.}{1998}]{vale98}
D'Odorico V., Cristiani S., D'Odorico S., Fontana A., Giallongo E., 
Shaver P., 1998, A\&A, 339, 678

\bibitem[\protect\citeauthoryear{D'Odorico et al.}{2006}]{vale06}
D'Odorico V., Viel M., Saitta F., Cristiani S., Bianchi S., Boyle B., Lopez S.,
Maza J., Outram P., 2006, MNRAS, 372, 1333

\bibitem[\protect\citeauthoryear{D'Odorico et al.}{2008}]{vale08}
D'Odorico V., Bruscoli M., Saitta F., Fontanot F., Viel M., Cristiani S., Monaco P., 2008, MNRAS, 379, 1727

\bibitem[\protect\citeauthoryear{Faucher-Gigu\`ere et al.}{2008}]{fg2008}
Faucher-Gigu\`ere C.-A., Prochaska J. X., Lidz A., Hernquist L.,
Zaldarriaga M., 2008, ApJ, 681, 831	

\bibitem[\protect\citeauthoryear{Francis et al.}{2001}]{francis01} 
Francis P. J., Catherine L. D., Matthew T. W., Michael J. W., Rachel L. W., 2001, PASA, 18, 221

\bibitem[\protect\citeauthoryear{Haardt \& Madau}{1996}]{hm96}
Haardt F., Madau P., 1996, ApJ, 461, 20

\bibitem[\protect\citeauthoryear{Janknecht et al.}{2006}]{janknecht06} 
Janknecht E., Reimers D., Lopez S., Tytler D., 2006, A\&A, 458, 427

\bibitem[\protect\citeauthoryear{Kaper et al.}{2009}]{kaper09} 
Kaper L., et al.\ 2009, in Proceedings ESO workshop on "Science with
the VLT in the E-ELT era", Ed. A. Moorwood, Springer Netherlands, pg. 319 

\bibitem[\protect\citeauthoryear{Kim, Cristiani \& D'Odorico}{Kim et al.}{2001}]{kim01} 
Kim T.-S., Cristiani S., D'Odorico S., 2001, A\&A, 373, 757

\bibitem[\protect\citeauthoryear{Kim et al.}{2002}]{kim02} 
Kim T.-S., Carswell R.F., Cristiani S., D'Odorico S., Giallongo E.,
2002, MNRAS, 335, 555 

\bibitem[\protect\citeauthoryear{Kim et al.}{2004}]{kim04}
Kim T. S., Viel M., Haehnelt M., Carswell R.F., Cristiani S., 2004, MNRAS, 347, 355

\bibitem[\protect\citeauthoryear{Kirkman et al.}{2005}]{kirkman05}
Kirkman D., Tytler D., Suzuki N., Melis C., Hollywood S., James K., So
G., Lubin D., Jena T., Norman M. L., Paschos P., 2005, MNRAS, 360, 1373

\bibitem[\protect\citeauthoryear{McDonald et al.}{2000}]{mcdonald00} 
McDonald P., Seljak U., Cen R., Bode P., Ostriker J. P., 2005, MNRAS, 360, 1471  
  
\bibitem[\protect\citeauthoryear{Ostriker, Bajtlik \& Duncan}{1988}]{ostriker88}
Ostriker J. P., Bajtlik S., Duncan R. C., 1988, ApJ, 327, 35
  
\bibitem[\protect\citeauthoryear{Rauch et al.}{2001}]{rauch01}
Rauch M., Sargent W. L. W., Barlow T. A., Carswell R. F., 2001, ApJ, 562, 76  
  
\bibitem[\protect\citeauthoryear{Rauch et al.}{2005}]{rauch05}
Rauch M., Becker G. D., Viel M., Sargent W. L. W., Smette A., Simcoe
R. A., Barlow T. A., Haehnelt M. G., 2005, ApJ, 632, 58

\bibitem[\protect\citeauthoryear{Rollinde et al.}{2003}]{rollinde03}
Rollinde E., Petitjean P., Pichon C., Colombi S., Aracil B., D'Odorico
V., Haehnelt M. G., 2003, MNRAS, 341, 1279

\bibitem[\protect\citeauthoryear{Savaglio et al.}{1999}]{savaglio99}
Savaglio S., Ferguson H. C., Brown T. M., Espey B. R., Sahu K. C., 
Baum S. A., Carollo C. M., Kaiser M. E., Stiavelli M., Williams R. E.,
Wilson J., 1999, ApJ, 515, L5

\bibitem[\protect\citeauthoryear{Scannapieco et al.}{2006}]{scannapieco06}
Scannapieco E., Pichon C., Aracil B., Petitjean P., Thacker R. J., Pogosyan
D., Bergeron J., Couchman H. M. P., 2006, MNRAS, 3365, 615

\bibitem[\protect\citeauthoryear{Schaye et al.}{2003}]{schaye03}
Schaye J., Aguirre A., Kim T. S., Theuns T., Rauch M., Sargent W. L. W., 2003, ApJ, 596, 768

\bibitem[\protect\citeauthoryear{Schlegel et al.}{2007}]{schlege07} 
Schlegel, D.J., et al.\ 2007, Bulletin of the American Astronomical Society, 38, 966 

\bibitem[\protect\citeauthoryear{Seljak et al.}{2005}]{seljak05} 
Seljak U., et al., 2005, PhysRevD, 71, 3511

\bibitem[\protect\citeauthoryear{Smette et al.}{1992}]{smette92}
Smette A., Surdej J., Shaver P. A., Foltz C. B., Chaffee F. H., Weymann R. J., 
Williams R. E., Magain P., 1992, ApJ, 389, 39

\bibitem[\protect\citeauthoryear{Spergel et al.}{2003}]{spergel03}
Spergel D. N., 2003,  ApJ Suppl. S., 148, 175

\bibitem[\protect\citeauthoryear{Springel, Yoshida \& White}{Springel et al.}{2001}]{springel2001}  
Springel V., Yoshida N., White S. D. M., 2001, NewA, 6, 79

\bibitem[\protect\citeauthoryear{Springel \& Hernquist}{2002}]{springel2002} 
Springel V., Hernquist L., 2002, MNRAS, 333, 649

\bibitem[\protect\citeauthoryear{Springel}{2005}]{springel2005} 
Springel V., 2005, MNRAS, 364, 1105

\bibitem[\protect\citeauthoryear{Tripp et al.}{1998}]{tripp98}
Tripp T. M., Limin Lu, Savage B. D., 1998, ApJ, 508, 200

\bibitem[\protect\citeauthoryear{Viel et al.}{2002}]{viel2002}
Viel M., Matarrese S., Mo H. J., Haehnelt M., Theuns T. 2002, MNRAS 329, 848

\bibitem[\protect\citeauthoryear{Viel, Haehnelt \& Springel}{Viel et al.}{2004}]{viel2004} 
Viel M., Haehnelt M. G., Springel V., 2004a, MNRAS, 354, 684

\bibitem[\protect\citeauthoryear{Viel, Colberg \& Kim}{Viel et al.}{2008}]{viel2008} 
Viel M., Colberg J. M. and Kim T.S., 2008, MNRAS, 386, 1285

\end{thebibliography}
\end{document}